\documentclass[12pt,a4paper,dvips]{article}
\usepackage{amssymb,amsmath,graphicx,a4p}
\usepackage{cite,mcite}
\usepackage{ifthen}
\usepackage{l3_title}
\usepackage{Lep}
\usepackage{epsfig,subfigure,higgs_input}
\journalname{Phys. Lett. B}
\date{July 10, 2001}
\preprint{2001-049}
\Lep{2}
\usepackage[nolists,noheads,nomarkers,tablesfirst]{endfloat}

\begin{document}
\begin{titlepage}
  
  \title{Standard Model Higgs Boson\\
 with the L3 Experiment at LEP}
  
  \author{L3 Collaboration}
%
%
  \begin{abstract}
    Final results of the search for the
    Standard Model Higgs boson are presented for the
    data collected by the L3 detector at LEP  
    at  centre-of-mass energies up to  about 209\GeV.  
    These data are compared with the expectations
    of Standard Model processes for Higgs boson masses up to 120 GeV.
A lower limit on the mass of the Standard Model Higgs boson of 112.0 GeV 
is set at the 95\% confidence level. 
The most significant high mass candidate is a \Hnn event. 
It has a reconstructed Higgs mass of 115\GeV and it was recorded at 
$\come=206.4\GeV$. 
  \end{abstract}
\vspace*{2.cm}
\begin{center}
{To be submitted to {\it Phys. Lett. B}
}
\end{center}
%
\end{titlepage}
%
%
\section{Introduction}
\label{sec:intro}
One of the most important goals of the L3 experiment at the LEP \epem 
collider was to find the Higgs boson. 
In the Standard Model this particle
is associated to the Higgs field~\cite{higgs_1}, 
expected to provide mass to 
all the observed elementary particles. The mass of the Higgs boson, \mH, 
is not predicted by the theory. Before the advent of LEP, there was no 
solid experimental information about the Higgs mass. 
The L3 experiment has carried out the search for the Higgs boson at LEP
in very large data samples collected at the Z 
resonance~\cite{l3_smh_1990} and at
ever increasing centre-of-mass energies and 
luminosities~\cite{l3_smh_172,l3_smh_183,l3_smh_189,l3_smh_202,l3_smh_2000_a} 
greatly extending the Higgs mass range investigated. 
 
A fit that includes L3 electroweak precision measurements results 
in an upper limit  on \mH of 133\GeV~\cite{l3ewhiggslimit} at the 
95\% confidence level. 
Previous L3 direct searches for the Standard Model Higgs boson   
excluded the mass range up to 107\GeV~\cite{l3_smh_202}. 
Similar results were also reported by other LEP experiments~\cite{al_smh_202}.
Results of the Standard Model Higgs search obtained shortly after the 
end of the LEP data taking  in the year 2000 were also published 
by L3~\cite{l3_smh_2000_a} and by the other 
LEP experiments~\cite{al_smh_2000_a}.

The Standard Model Higgs boson is produced at LEP 
mainly via the Higgs-strahlung process
  \epemtoSMHZ . 
The processes of \WW and \ZZ fusion contribute, with smaller rate,  
to the Higgs
production in the \Hnn and \Hee channels, respectively. 
The largest sources of background are 
four-fermion final states from \W  and \Z pair production, as well as
quark pair production \epemtoqqg.

In this letter, the final results of the Standard Model Higgs search 
performed on the data collected by L3 at LEP at a 
centre-of-mass energy, \come, up to about 
209\GeV are reported. 
These results 
include the full luminosity collected in the year 2000 
and the corrected LEP beam energies. 
In the year 2000 LEP was run at several values of
\come.
The slight beam energy adjustments significantly affect the 
signal expectation at the highest Higgs masses, close to the 
kinematic limit for HZ production, $\mH=\come-\mZ$, where \mZ is the 
Z boson mass. The effect of the Z width
in the Higgs mass reconstruction close to the HZ kinematic limit is also 
taken into account. 
Final calibrations of all subdetectors are applied.
The signal and background
expectations are evaluated on a finer grid of \come values, 
with larger samples of 
simulated events, thus reducing statistical and systematic uncertainties.  
Therefore, the results reported in this letter are affected by total  
uncertainties smaller than in Reference~\citen{l3_smh_2000_a}.


%
\section{Data and Monte Carlo samples}
\label{sec:datamc}
The data were collected by the L3
detector~\cite{l3_1990_1}
at LEP during the year 2000 at several centre-of-mass energies.
The total collected luminosity amounts to 217.3\pb.
The  data are grouped into seven samples corresponding to average 
centre-of-mass energies between 202.8\GeV and 208.6\GeV.   
The integrated luminosities corresponding to these
samples are given in Table~\ref{table:lumi}.

The Higgs production cross sections and branching ratios are calculated
using the HZHA generator~\cite{janotlepii_v2}.
Efficiencies are determined using Monte Carlo samples of Higgs events, 
generated with PYTHIA~\cite{PYTHIA}.
Since the Higgs production cross sections and efficiencies depend 
strongly on \come, in particular for \mH close to the HZ kinematic limit, 
samples of  Higgs events are simulated at each centre-of-mass energy shown in 
Table~\ref{table:lumi}. Higgs events are generated with \mH between 
105 and 120\GeV, in steps of 1\GeV. For each mass and each search 
channel, between 2000 and  10000 events are generated.
 
The Standard Model background estimates
rely on  the following Monte Carlo programs: 
KK2f~\cite{KK2f} for \epemtoqqg,
KORALW~\cite{KORALW} for \epemtoWW, 
PHOJET~\cite{PHOJET} for two-photon processes (\epemtoeeqq) and  
EXCALIBUR~\cite{EXCALIBUR} for other four-fermion final states.
The number of simulated
events for the dominant backgrounds is at least 100
times the number of events expected for such processes.
   
The response of the L3 detector is simulated using the 
GEANT program~\cite{xgeant}, taking into account the effects of multiple
scattering, energy loss and showering in the detector. Hadronic
interactions in the detector are modelled using the GHEISHA
program~\cite{xgheisha}. 
Time dependent detector inefficiencies,
as monitored during the data taking period, are also simulated.


%
\section{Analysis procedures}
The search for the Standard Model  Higgs boson is based on the study of 
four distinct event topologies:
\SMHZtoqqqq,  
\SMHZtoqqnn,
\SMHZtoqqll ($\ell=\mathrm{e},\mu,\tau$) 
and \SMHZtottqq. \\
In the following they are denoted as \Hqq, \Hnn, \Hll and  \ttqq, respectively.
With the exception of the \SMHZtottqq 
decay mode,  all the analyses are optimised
for the $\bigH\!\rightarrow\!\bbbar$ decay. 
This mode represents about 80\% of
the Higgs branching fraction in the mass range of interest. 

\begin{figure}[hp]
\mbox{\epsfig{figure=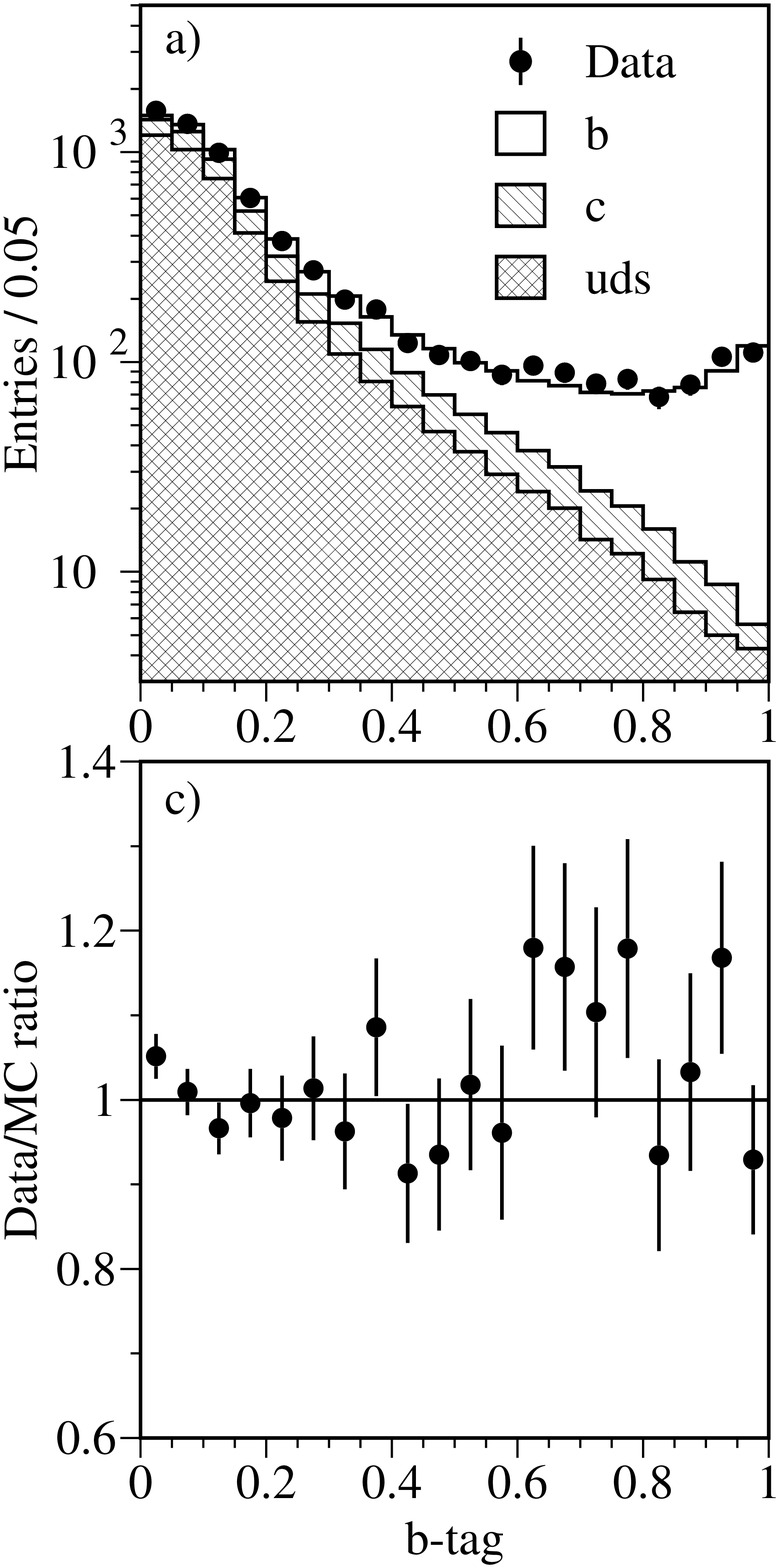,width=0.5\textwidth}%
      \epsfig{figure=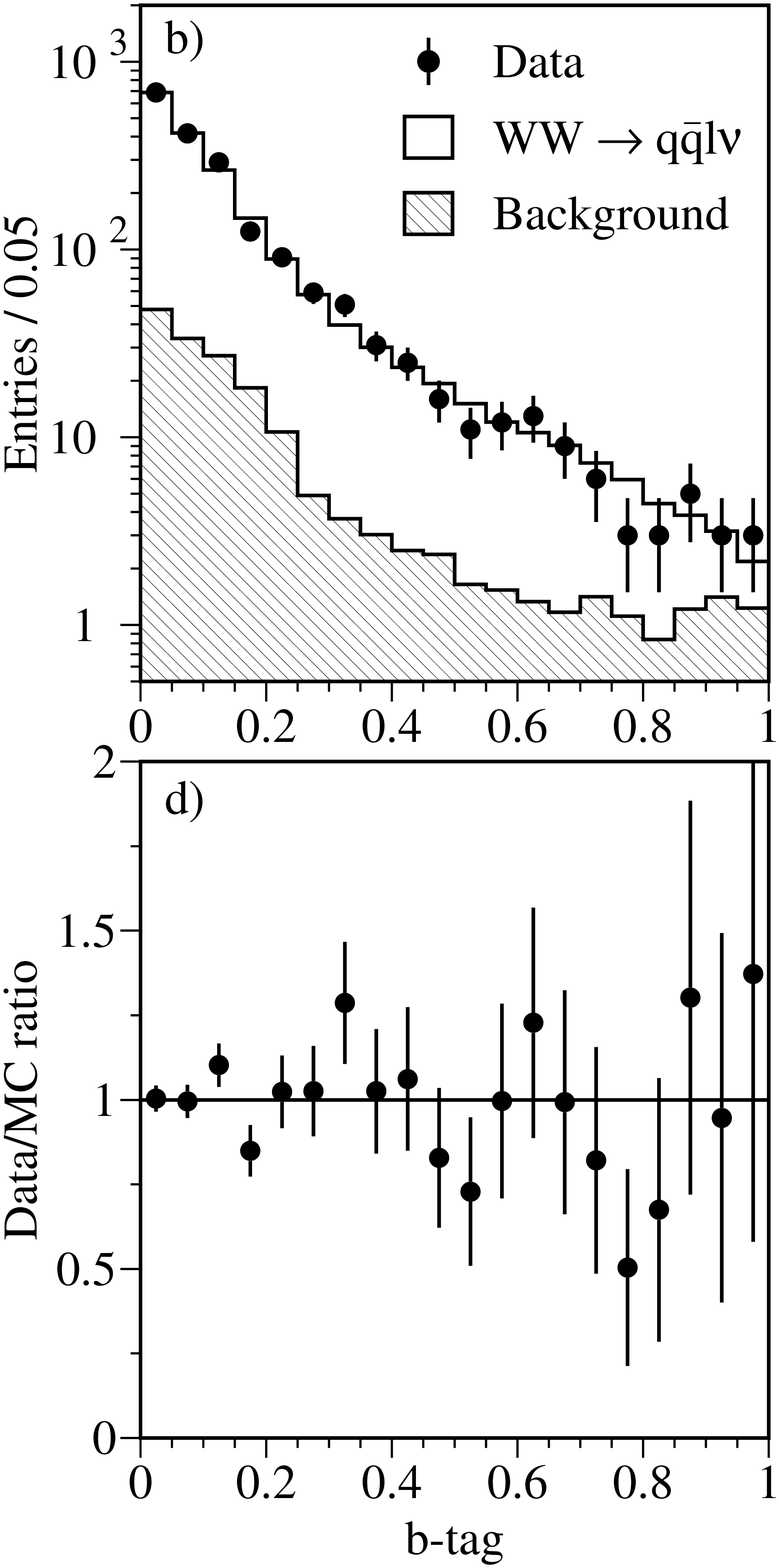,width=0.5\textwidth}}
\caption[]{\label{fig:btagcontrol}  Distribution  of the neural network jet b-tag  variable
           in a  sample  of a)  \epemtoqqg  and b)  \WWtoqqlnu
           events  selected from the  high-energy  data collected in the
           year  2000.  Two entries per event contribute to  the distributions. The  data  are  compared  to the  simulation  of
           Standard Model  processes.  The bin-by-bin  ratio of the data
           to the simulated events is displayed in c) and d).}
\end{figure}

All the search channels are analysed in three stages.
First, a high multiplicity hadronic event selection is applied
to reduce the large background from two-photon processes, 
while preserving most of the Higgs signal.  
In a second stage,  topological and kinematical variables
together with b-tag variables 
are either used  to construct an event likelihood 
or fed into a neural network, to further discriminate between 
signal and background events.
A b-tag variable  is calculated for each hadronic jet using a neural
network~\cite{l3_smh_172} which exploits three-dimensional decay lengths,
properties of semileptonic b decays and jet-shape variables.  
The tracking and b-tagging performance in the Monte Carlo simulation 
are tuned using 4\pb of calibration data collected at $\come\sim\mZ$ in the year 2000.
The b-tagging performance for the high-energy data  
is verified with samples of \epemtoqqg events. 
The efficiency for tagging light flavoured hadrons is verified with 
\WWtoqqlnu events. 
The agreement of data with the simulation of Standard Model processes  
in the jet b-tag variable based on neural network is shown in Figure~\ref{fig:btagcontrol} for the 
\epemtoqqg and the \WWtoqqlnu events. 
The expectation from the Standard Model Monte Carlo describes the data within 
the statistical uncertainty. 

The neural network b-tag variables are combined into an event b-tag variable.
First,
 the probability is calculated for each jet to be compatible with the distribution
for light quarks determined from Monte Carlo. Then, the event b-tag variable is defined as
the negative log-likelihood of these probabilities.

The last part of the analysis is the construction of 
a final discriminant for each topology.
It is built from a combination
of the event likelihood, or the neural network output, 
with the reconstructed Higgs mass.
For each Higgs mass hypothesis, 
the final discriminants are computed for
the data and  for the expected background and signal. 
The distributions of the final discriminants
are then used to calculate the likelihood ratio, $Q$, as a function of \mH. 
This is the  ratio of the probability of 
observing the data in the presence of both the signal and the background,
``signal+background'' hypothesis, to the probability
of observing the data in the presence of only the background, 
``background-only'' hypothesis. 
The quantity used to evaluate the compatibility of the data 
with the signal is the log-likelihood ratio defined by~\cite{lephwg_2000}: 
\[
-2\ln{Q}\equiv 2\sum_i [s_i - n_i \ln{(1+s_i/b_i)}]\ .
\]
In this expression, $i$ indicates the $i$-th bin of the 
final discriminant of each channel and at each \come; $n_i$, $s_i$ and 
$b_i$ indicate respectively the number of observed events, 
the expected Higgs signal and the Standard Model background, 
in the $i$-th bin. 
Each event in the sum has a weight $\ln{(1+s/b)}$ 
which depends on the signal-to-background ratio, $s/b$, in the bin 
where it is found. 
This weight depends on the Higgs mass hypothesis.    
For each given \mH, the value of the log-likelihood ratio
in the data is compared to the expected distributions of $-2\ln{Q}$  
in a large number of simulated experiments under  
the ``background-only'' and the ``signal+background'' hypotheses.
The results for each search channel are then presented in terms of 
$-2\ln{Q}$ for the data compared to the expected median values for 
the two hypotheses, as a function of \mH.


%
\subsection{The {\boldmath\Hqq} analysis}


The \Hqq analysis aims to select and study events with four jets, 
two of which contain b hadrons, while the other two must be
consistent with the decay of a \Z boson. 
Background from Standard Model  processes comes mainly from \qqbar final states 
with hard gluon radiation, \WW and \ZZ events, especially those 
where one of the \Z bosons decays into b quarks. 

After a high multiplicity hadronic preselection,  the  events are forced into
four jets with the DURHAM algorithm~\cite{DURHAM}
and a kinematic fit requiring four-momentum conservation is performed. 
Several discriminating variables,  ${x_i}$, are combined into a single
likelihood which is then used to select the final sample.
In this combination, each final state is considered as an event class 
$j$ ($j$=HZ, ZZ, WW, $\qqbar$).
For each class,  probability density functions ${f^{j}(x_i)}$ are derived from
Monte Carlo.
The probability for an event to belong to the event class $j$,
based on the value of the variable ${x_i}$, is  
defined as
$ p^{j}(x_i)=f^{j}(x_i)/\sum_k f^{k}(x_i)$,        
where $k$ runs over all classes.

\begin{figure}[hp]
\mbox{\epsfig{figure=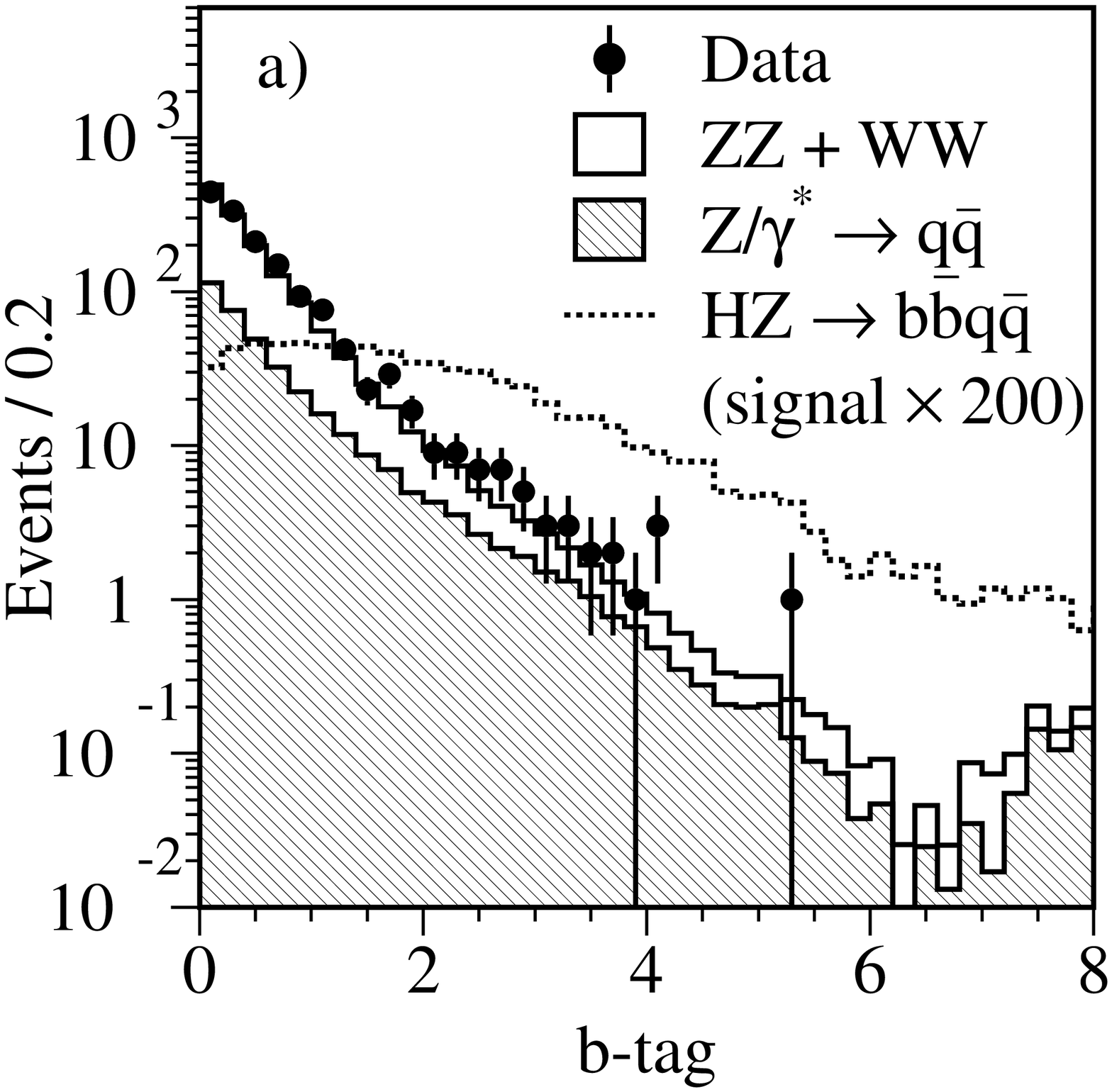,width=0.5\textwidth}%
      \epsfig{figure=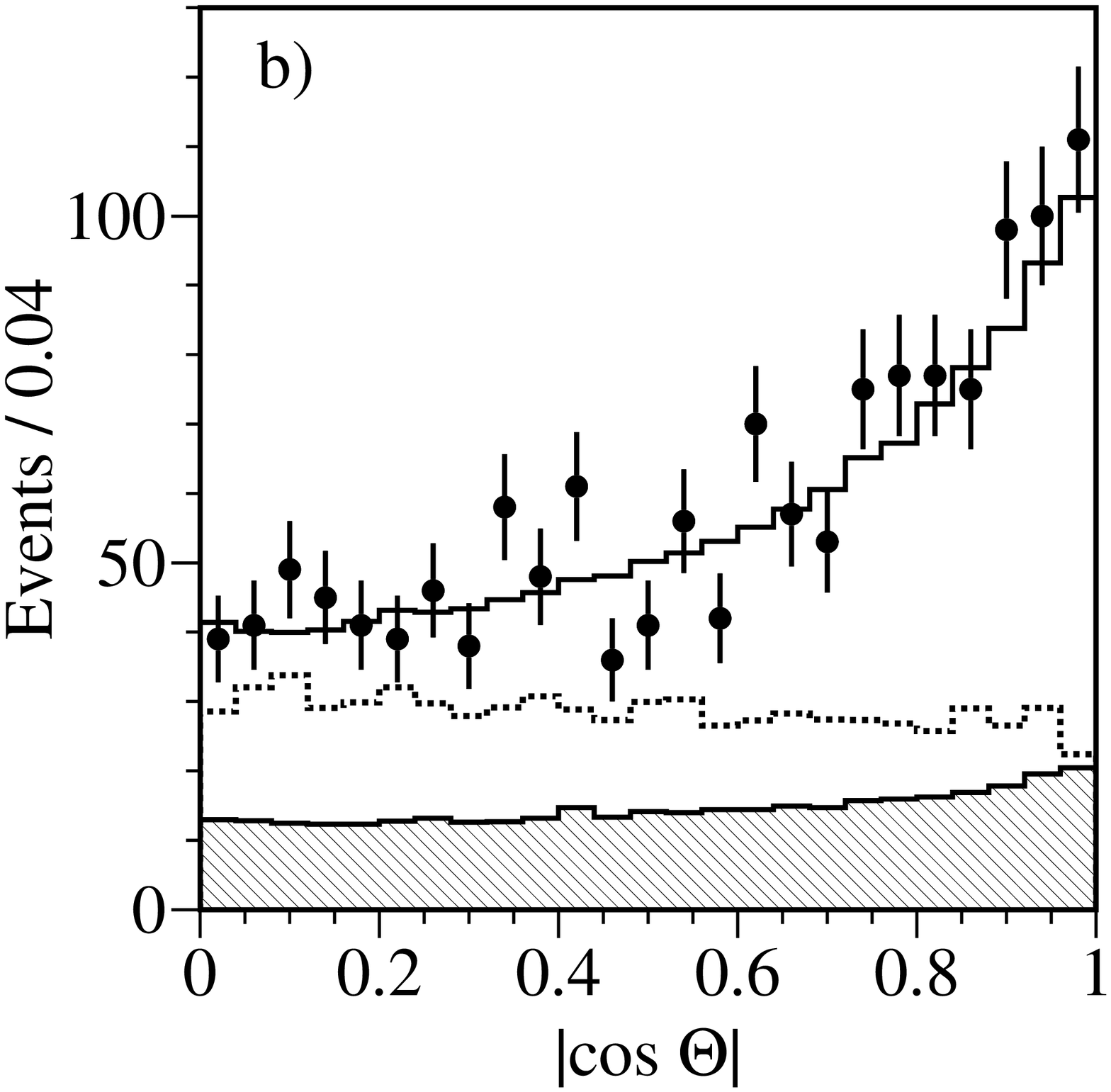,width=0.5\textwidth}}
\mbox{\epsfig{figure=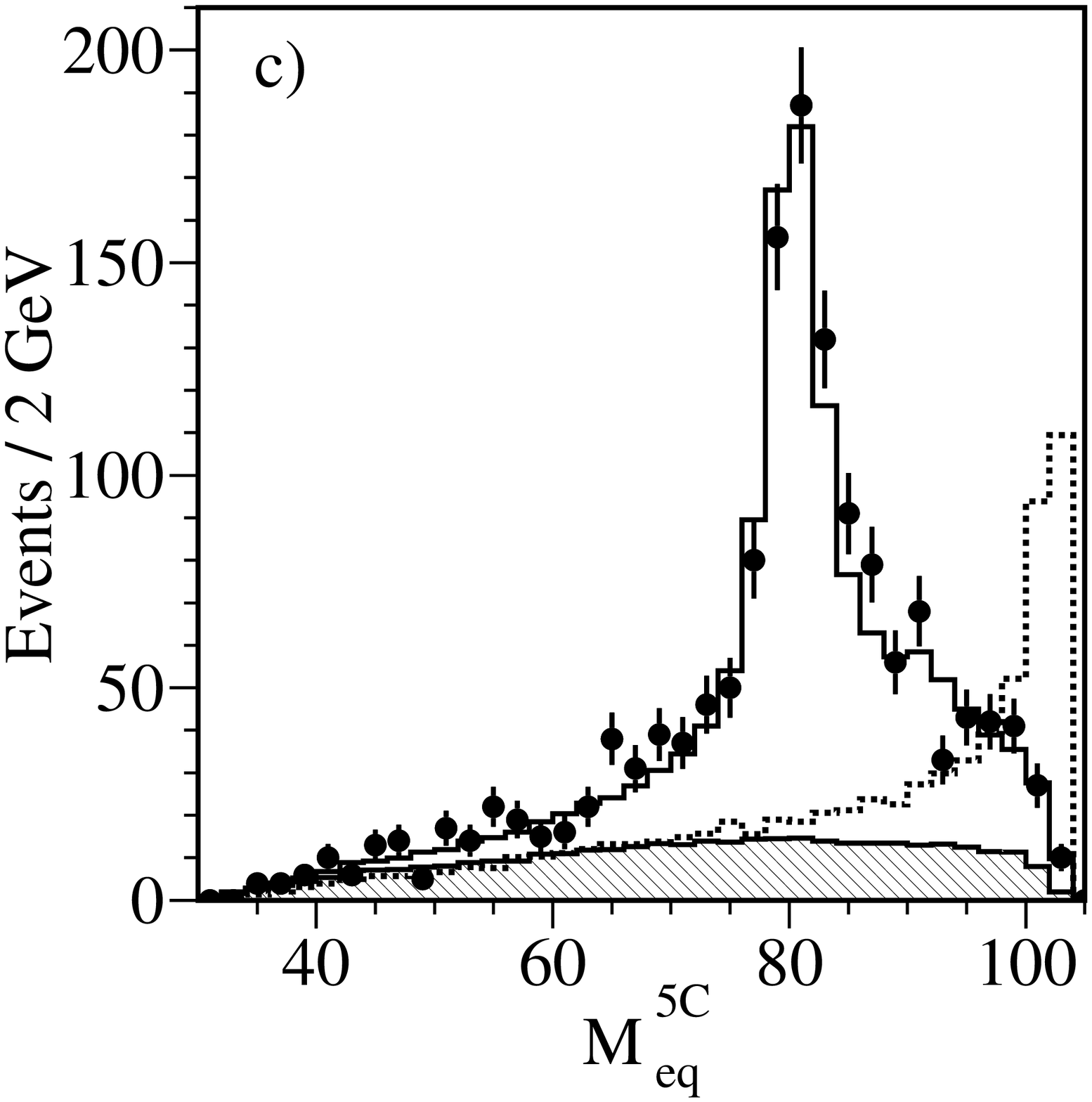,width=0.5\textwidth}%
      \epsfig{figure=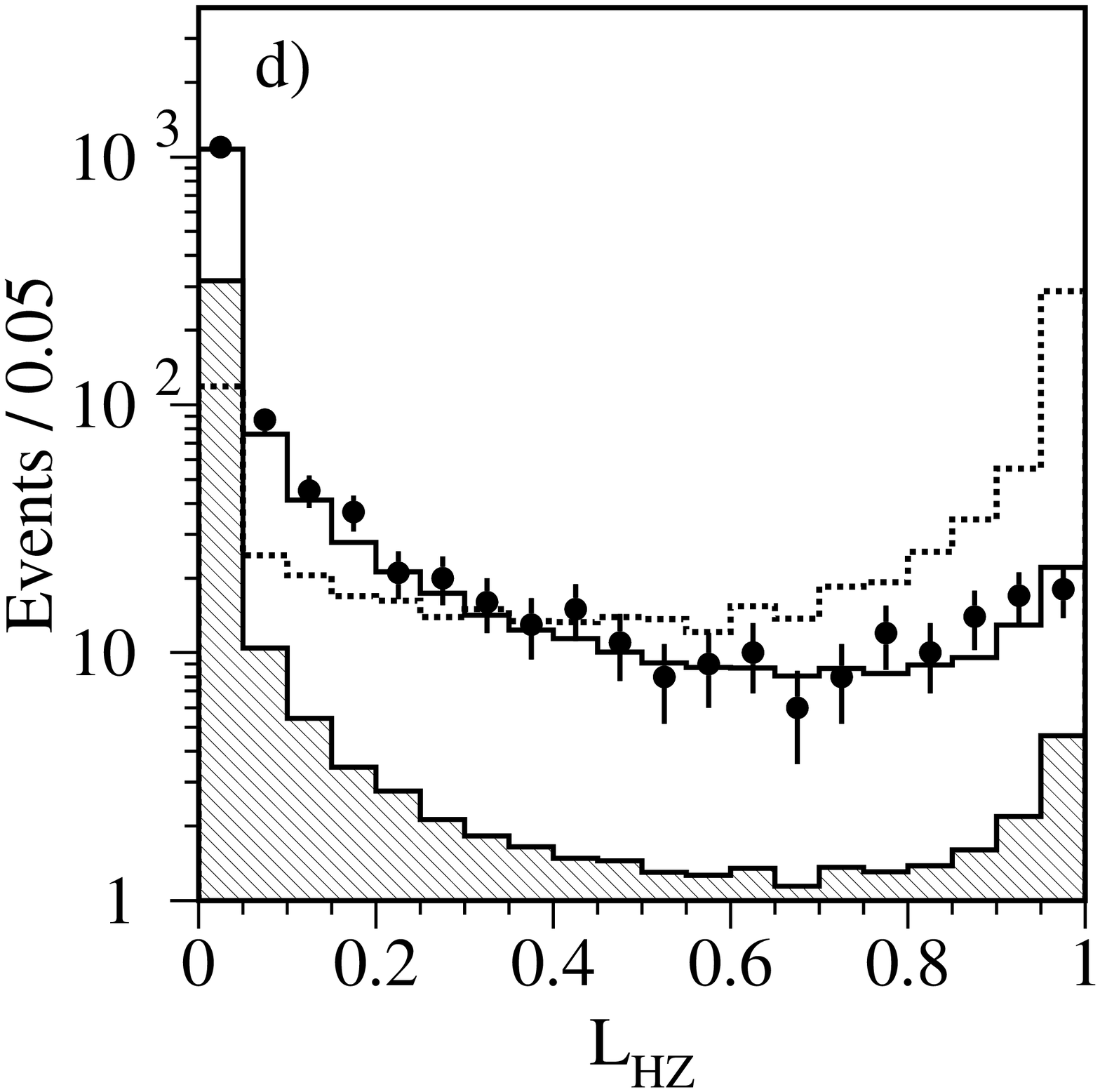,width=0.5\textwidth}}
\caption[]{\label{fig:qqqqvari}  Distribution  of a) the event  b-tag,  b) the
           cosine of the boson production angle, c) the mass from the 5C
           equal-mass fit and d) the likelihood for the events  selected
           in the \Hqq  search  channel.  The points  correspond  to the
           data  collected  at $\rts > 206 \GeV$.  The open and  hatched
           histograms are the expected  backgrounds  from Standard Model
           processes.  The dashed line is the distribution  expected for
           a 115\GeV Higgs signal, multiplied by a factor of 200.}
\end{figure}

The individual probabilities are combined into a likelihood:
$L_{\rm HZ}=\prod_i p^{\rm HZ}(x_i)/\sum_k\prod_i p^{k}(x_i)$, 
where $i$ runs over all variables considered and $k$ over all event classes.
Ten variables are used to calculate the likelihood. They are:
\begin{itemize}
\item the number of tracks, 
\item the event b-tag,
\item the maximum energy difference between any two jets, 
\item the minimum jet energy, 
\item the parameter of the DURHAM algorithm for which the 
    event is resolved from  three jets
    into  four jets,  
\item the maximal triple-jet boost,
    defined as the maximum three-jet boost obtained
    from the four possibilities to construct a 
    one-jet against three-jet configuration in a four-jet event,
\item the minimum opening angle between any two jets, 
\item the event sphericity, 
\item the mass from a 5C kinematic fit
imposing energy and momentum conservation and 
equal dijet masses, ${M_{eq}^{5C}}$, 
\item the  absolute value of the cosine of the production polar angle,
$|\cos{\Theta}|$,
assuming the production of a pair of bosons.
\end{itemize}
The distributions of the event b-tag, 
${M_{eq}^{5C}}$,  $|\cos{\Theta}|$ and ${L_{\rm HZ}}$
for the events selected in the \Hqq search channel with $\come>206\GeV$, 
compared to the 
expectation for Standard Model processes, 
are shown in Figure~\ref{fig:qqqqvari}.

Events are selected into the final sample if the value of 
${L_{\rm HZ}}$ exceeds a threshold 
optimised for each centre-of-mass energy and each
Higgs mass hypothesis. 
In addition, the compatibility 
of each event with a Higgs mass hypothesis \mH
is tested by the variable 
$ \chi^2_{\rm HZ}=\big( \Sigma_i - (\mH+\mZ) \big)^2/\sigma_{\Sigma_{\rm HZ}}^2
+ \big( \Delta_i - |\mH - \mZ| \big)^2/\sigma_{\Delta_{\rm HZ}}^2
$. In this expression, $\Sigma_i$ and $\Delta_i$ are the dijet mass sum and 
dijet mass difference, respectively, for the $i$-th of the three 
possible jet pairing combinations, 
while $\sigma_{\Sigma_{\rm HZ}}$ and 
$\sigma_{\Delta_{\rm HZ}}$ are the corresponding resolutions for Higgs events. 
The jet pairing with the best $\chi^2$ is chosen.
Finally, only events with the $\chi^2$ probability above 0.01  are selected.
As an example, for $\mH=110\GeV$, 179 events are selected in the data
with 172 expected from background processes and 12.8 events expected 
from the Higgs signal; 
for  $\mH=115\GeV$, 149 events are observed with 142 from background and 
3.2 from the Higgs signal. 

For these events a final discriminant is constructed. 
At first, the events are classified into three categories depending 
on the values of the b-tag of the two jets assigned to 
the Higgs boson.
The first category contains events where none of these jets 
has the highest value of the b-tag among the four jets of the event.
The second category is composed of events
where one of these jets has the highest
b-tag value. 
The third category contains
events where the two jets assigned to the Higgs boson have the highest
b-tag values. The  $\mathrm{\chi^2_{HZ}}$ probability, 
the b-tag values of the individual jets and
the event category are then combined into the final discriminant.


%
\subsection{The \boldmath\Hnn analysis}
The \Hnn search is based on the selection of
events with two  jets containing b hadrons, with large missing energy 
and missing mass consistent with \mZ. 
A neural network is used for the \Hnn analysis,
very similar to the one previously 
reported~\cite{l3_smh_2000_a,l3_smh_202}. 
However, tighter cuts on radiative photons and on the jet polar angle 
are applied to reduce the \epemtoqqg background and ensure 
the best jet energy and b-tag measurements. The signal efficiency is 
slightly reduced by a few percent relative but the search 
performance is not significantly modified.
In addition, the neural network is trained with the final, 
high statistics, signal and background Monte Carlo samples, 
at each \come value, to maximise the sensitivity of the analysis.

In the first step of the analysis, high multiplicity hadronic events 
are selected and forced into two jets using the DURHAM algorithm.
The dijet invariant mass must exceed 40\GeV.  
These requirements reduce contributions from two-photon
interactions, while retaining a significant fraction of hadronic events
from \epemtoqqg and \W-pair production.  These  backgrounds
are then reduced by requiring the visible mass to be less than
140\GeV and the mass recoiling against the hadronic system to lie
between 50\GeV and 130\GeV.

\begin{figure}[hp]
\mbox{\epsfig{figure=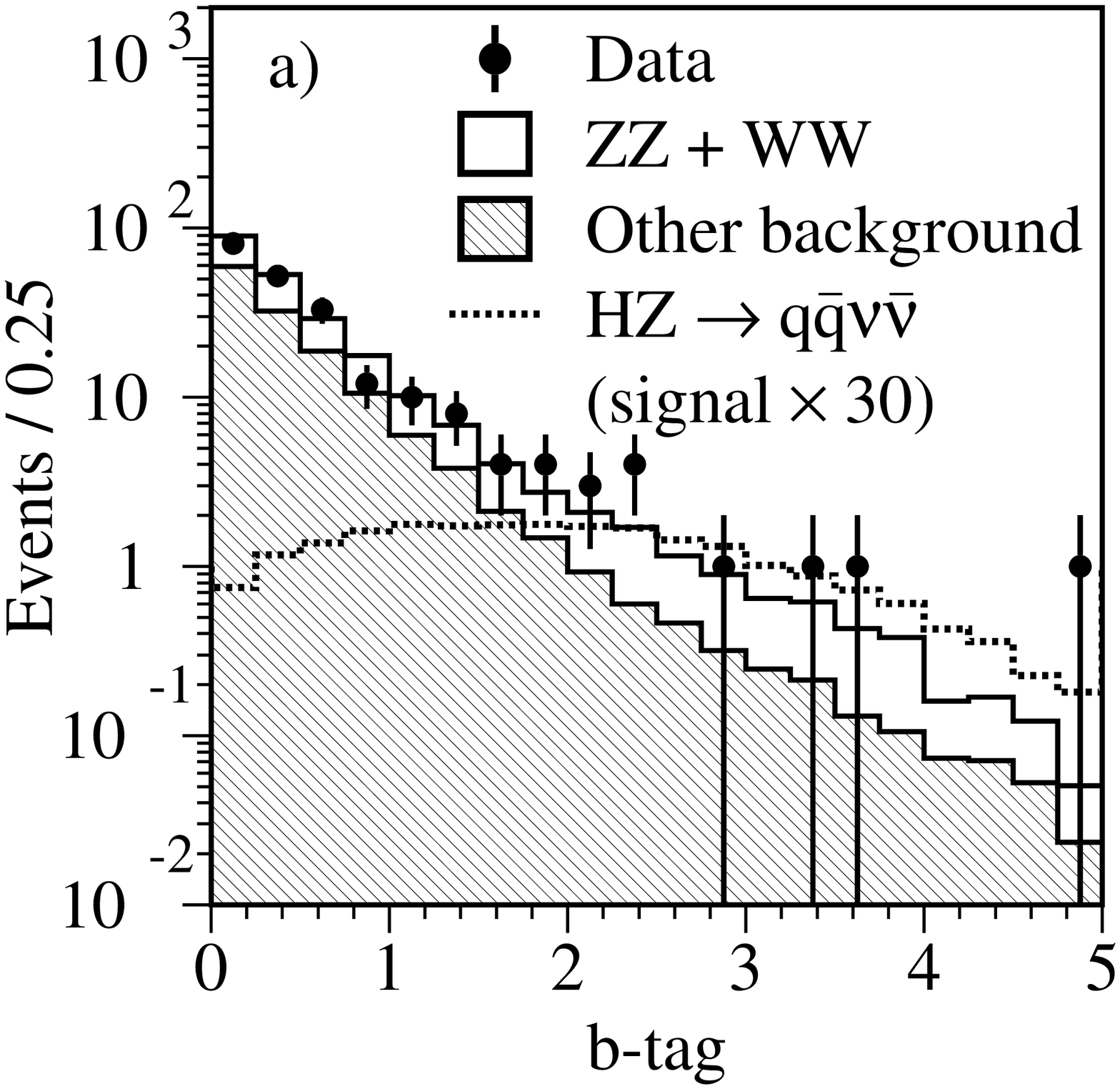,width=0.5\textwidth}%
      \epsfig{figure=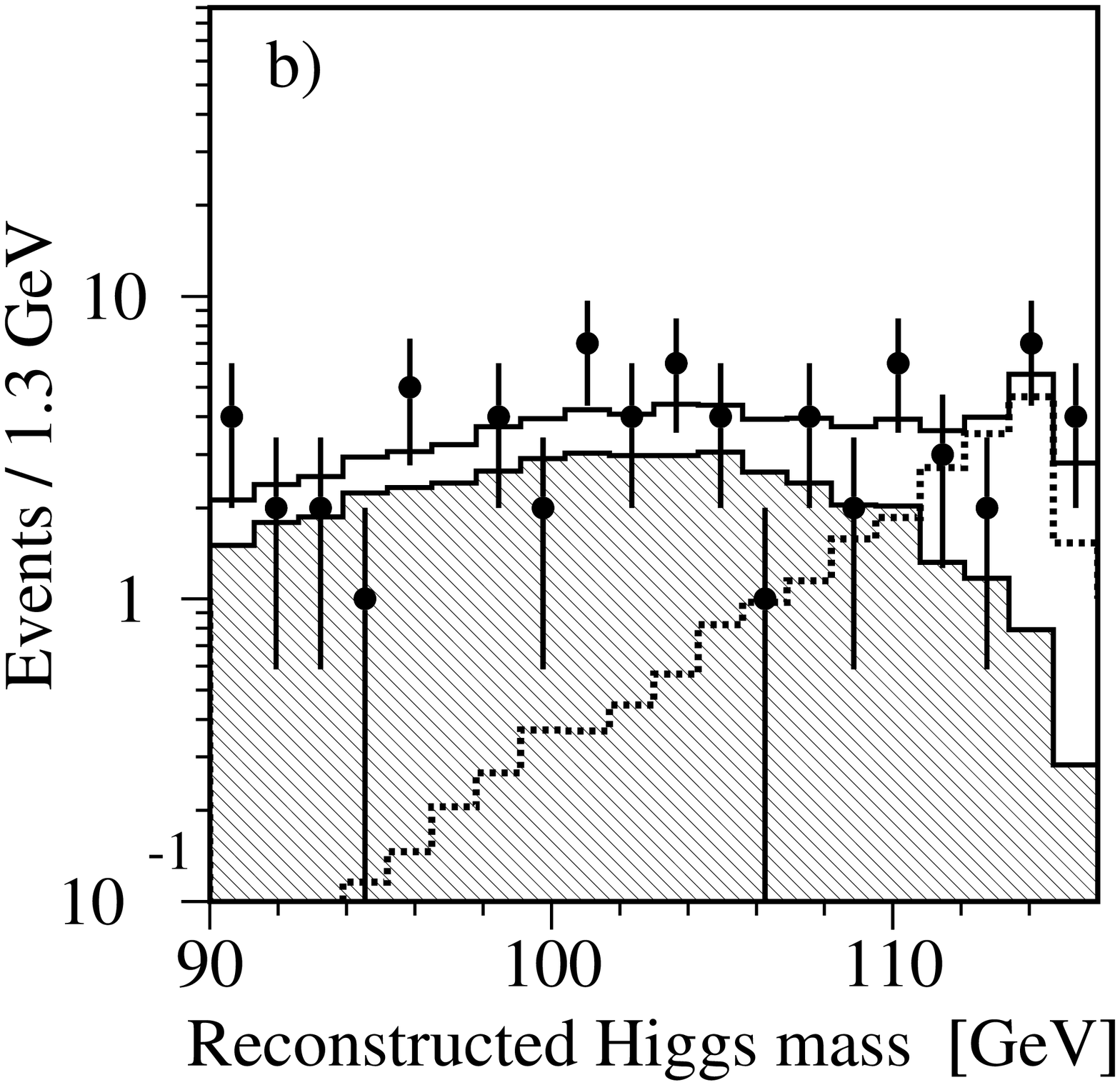,width=0.5\textwidth}}
\mbox{\epsfig{figure=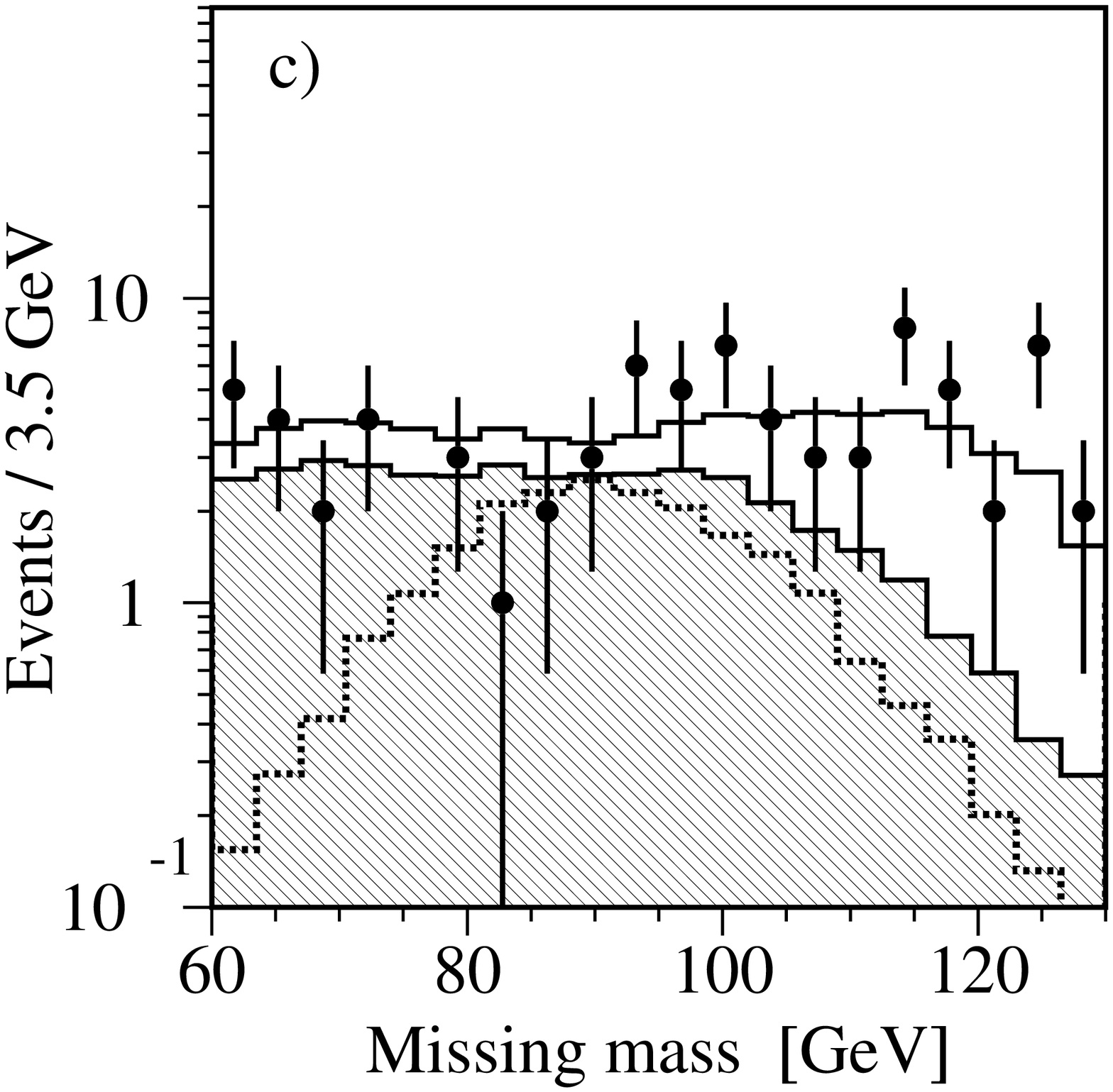,width=0.5\textwidth}%
      \epsfig{figure=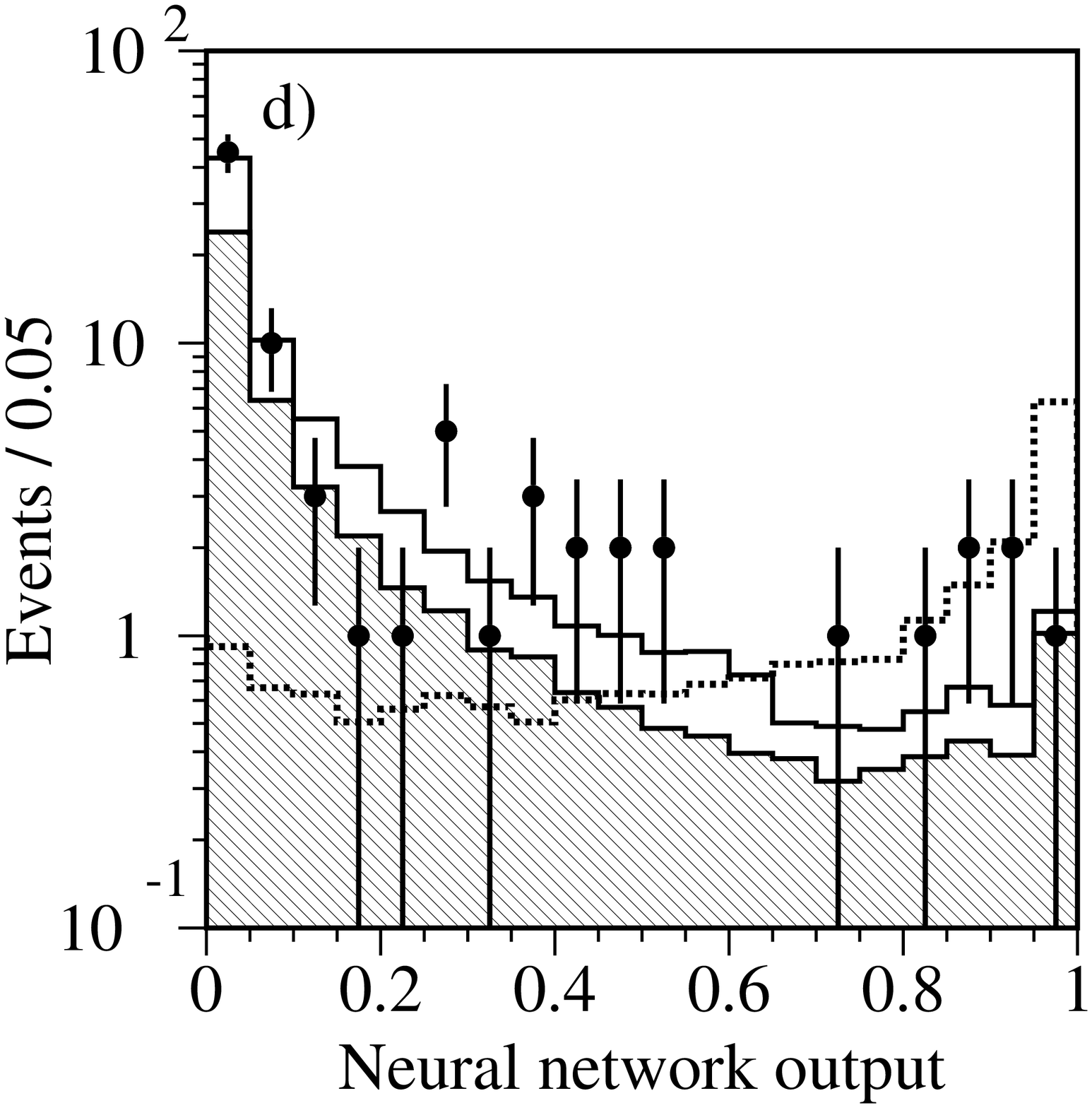,width=0.5\textwidth}}

\caption[]{\label{fig:qqnnvari}  Distribution  of a) the event b-tag,  b) the
           reconstructed  Higgs  mass,  c) the  missing  mass and d) the
           neural  network  output, for the events  selected in the \Hnn
           search  channel.  The points  represent the data collected at
           $\rts > 206 \GeV$.  The open and hatched  histograms  are the
           expected  backgrounds.  The dashed line is the expected Higgs
           signal with $\mH=115\GeV$, multiplied by a factor of 30.}
\end{figure}

Events from \epemtoqqg are further suppressed by requiring
the longitudinal missing energy to be less than $0.6\rts$ and 
the missing momentum vector to be at least $16^\circ$ away from the 
beam axis.  
The energy in the forward luminosity calorimeter is required to be below 
20\GeV. The acollinearity is required to be smaller than 65$^\circ$.   
The distribution of the event b-tag after the above cuts is shown in 
Figure~\ref{fig:qqnnvari}~a.
A loose cut requiring the event b-tag to be larger than 0.5 
is then applied, without further loss of signal efficiency.
After this set of cuts, there are 
123 events in the data, while 130 are expected from background 
processes with 4.3 and 1.3 events expected for $\mH=110\GeV$ and 115\GeV, 
respectively.

A kinematic fit imposing
four-momentum conservation and requiring the missing mass to be consistent 
with \mZ is
performed to compute the reconstructed Higgs mass from the two jets. 
The output of a mass independent neural network~\cite{l3_smh_183} 
is then combined with the reconstructed Higgs mass 
to build the final discriminant.
The distributions of  
   the reconstructed Higgs mass,  
   the missing mass 
   and the neural network output  
for the events selected in the \Hnn search channel with $\come>206\GeV$, 
compared to the 
expectation for Standard Model processes, 
are shown in Figure~\ref{fig:qqnnvari}. 
General agreement between the data and the expected 
contributions from Standard Model processes  is 
observed in all the distributions. 


%
\subsection{The \boldmath\Hll and \ttqq analyses}
The signatures for the \Hee and \Hmm processes 
are a pair of high energy electrons or muons with an invariant 
mass compatible with \mZ and two hadronic jets with b quark content.
In \Htt events the tau pair invariant mass must also be compatible
with \mZ. For these events, the mass resolution is worse than in the 
other \Hll channels due to 
the missing neutrinos from 
the tau decays. Events with Higgs decaying into tau leptons, \ttqq, have 
similar signature to the \Htt events, with the difference that 
the hadronic jet mass must be compatible with \mZ and that the b-tag content
of the event is reduced. 

The analyses are very similar to those described in 
Reference~\citen{l3_smh_202}.
The selections require high multiplicity events.
In the \Hee and \Hmm analyses 
two well identified electrons or muons are also required. 
In the tau analyses,  tau leptons are identified either by their
decay into electrons or muons, or as an isolated low-multiplicity jet
with one or three tracks and unit charge.
The identified leptons must have a large opening angle and must be
well isolated from the hadronic jets. 
For all \Hll selections, the invariant mass of the leptons after a kinematic
fit imposing four-momentum conservation must be consistent with \mZ within 
a mass range depending on the dilepton mass resolution.
In the \ttqq selection the mass of the two hadronic 
jets after kinematic fit must be consistent with \mZ.

After the \Hll selection, 18 events are observed with 16.7 
expected from background
processes and 1.7 or 0.32 signal events expected for \mH=110\GeV or 115\GeV,
respectively. After the \ttqq selection, 8 events are observed with 7.8 
expected from background and 0.66 or 0.15 signal events expected 
for \mH=110\GeV or 115\GeV, respectively.  

\begin{figure}[hp]
\mbox{\epsfig{figure=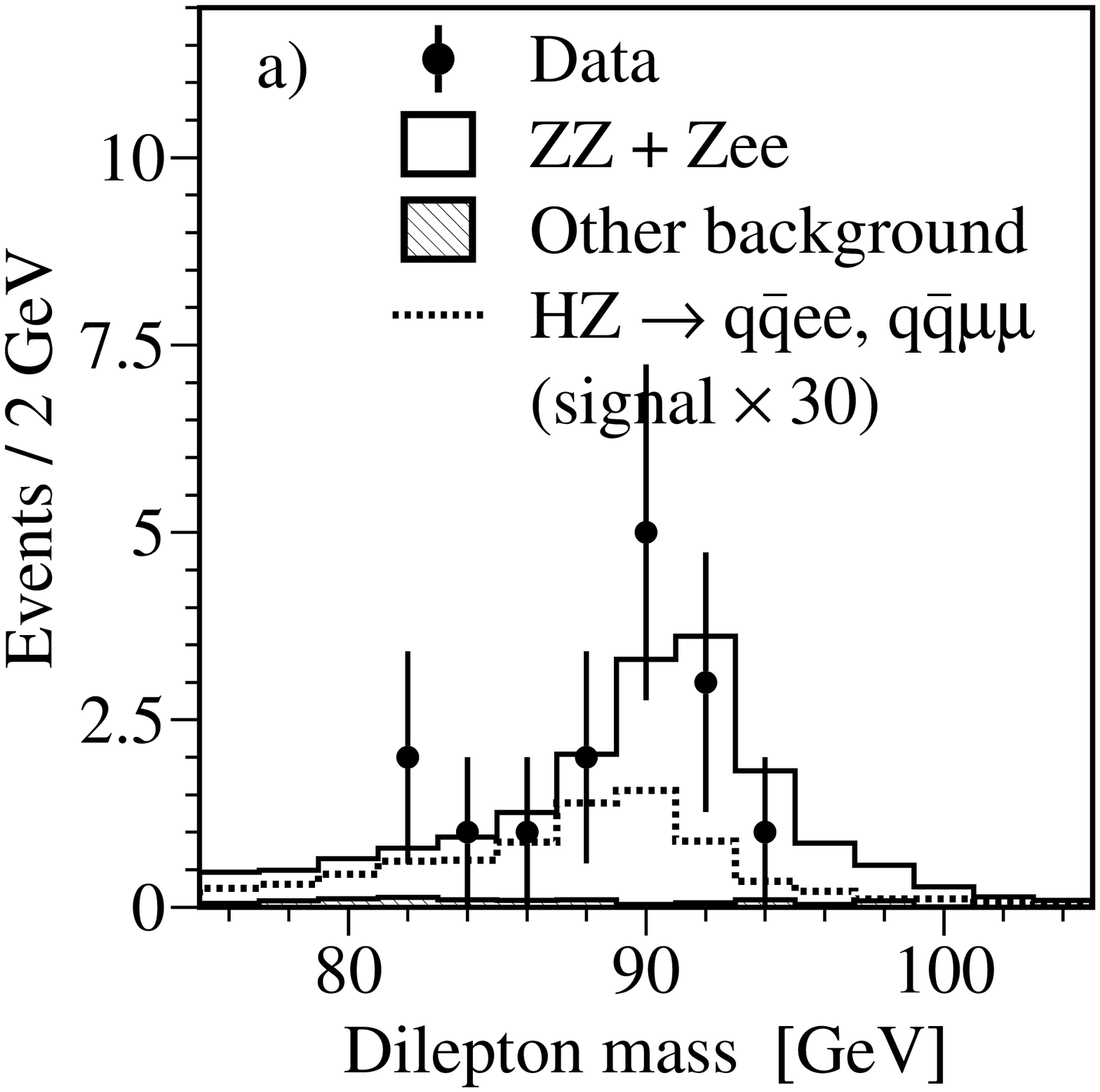,width=0.5\textwidth}%
      \epsfig{figure=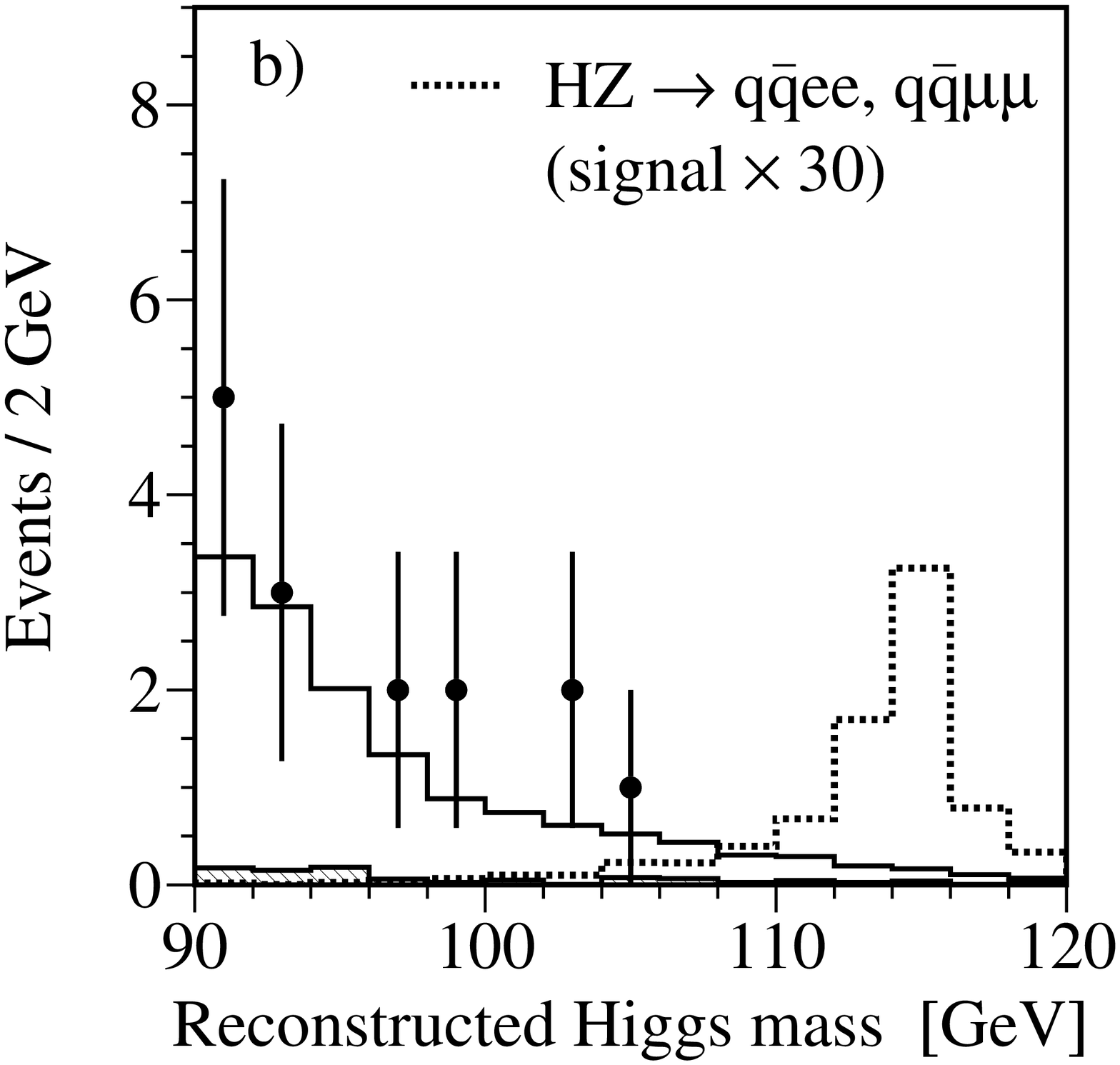,width=0.5\textwidth}}
\mbox{\epsfig{figure=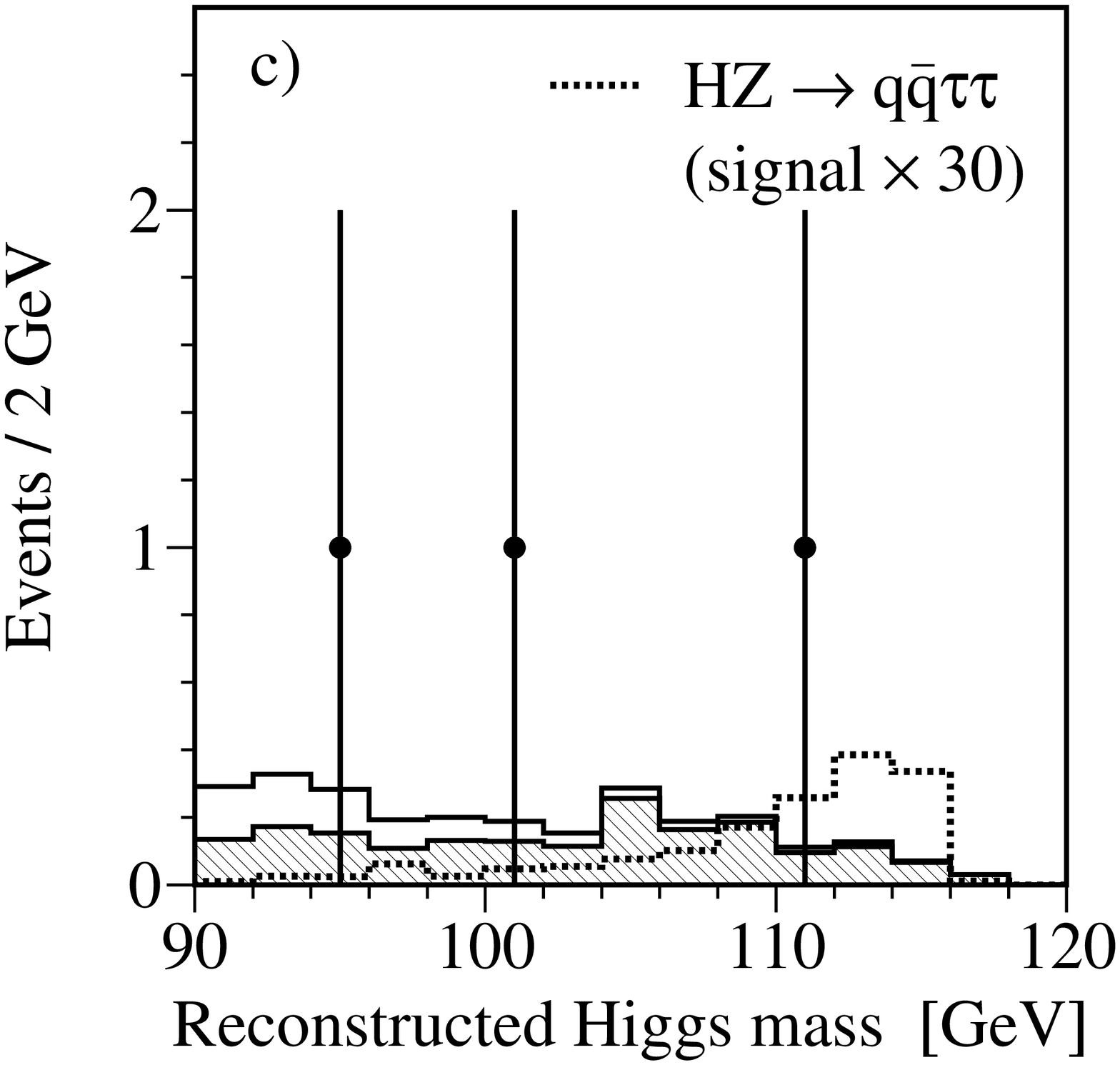,width=0.5\textwidth}%
      \epsfig{figure=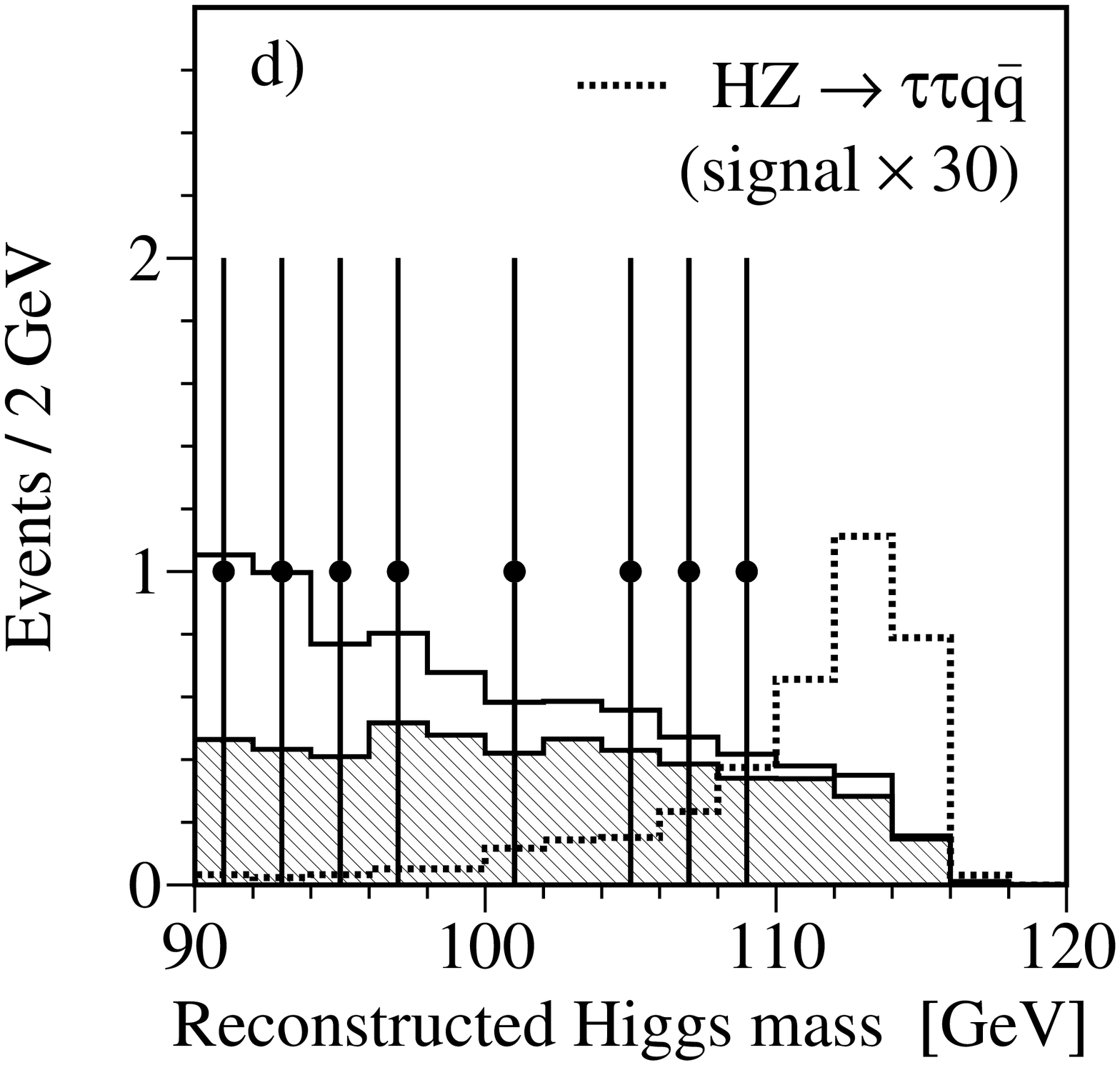,width=0.5\textwidth}}

\caption[]{\label{fig:qqllvari}  Distributions  of a) the dilepton  mass
           and the reconstructed  Higgs mass in the b) \Hee and \Hmm, c)
           \Htt and d) \ttqq  channels.  The points are the data and the
           open and hatched  histograms  the expected  backgrounds.  The
           dashed  line  is  the  expected  Higgs  signal  with  $\mH  =
           115\GeV$, multiplied by a factor of 30, in each channel.}
\end{figure}

The distributions of the dilepton mass and 
   the reconstructed Higgs mass in the \Hee and \Hmm channels
   are shown in Figure~\ref{fig:qqllvari}a and \ref{fig:qqllvari}b.  
The distributions of the reconstructed Higgs mass in the \Htt and 
\ttqq channels are shown in Figure~\ref{fig:qqllvari}c and 
\ref{fig:qqllvari}d, respectively.

In the \Hll selections,  the dijet mass after the fit is combined with the 
b-tag values  of the two jets, to form 
the final discriminant. For the \ttqq selection, the mass
of the tau pair, calculated by constraining the invariant mass of
the two other jets to \mZ, is used as the final discriminant.  


%
\section{Results}

\begin{figure}[hp]
\mbox{\epsfig{figure=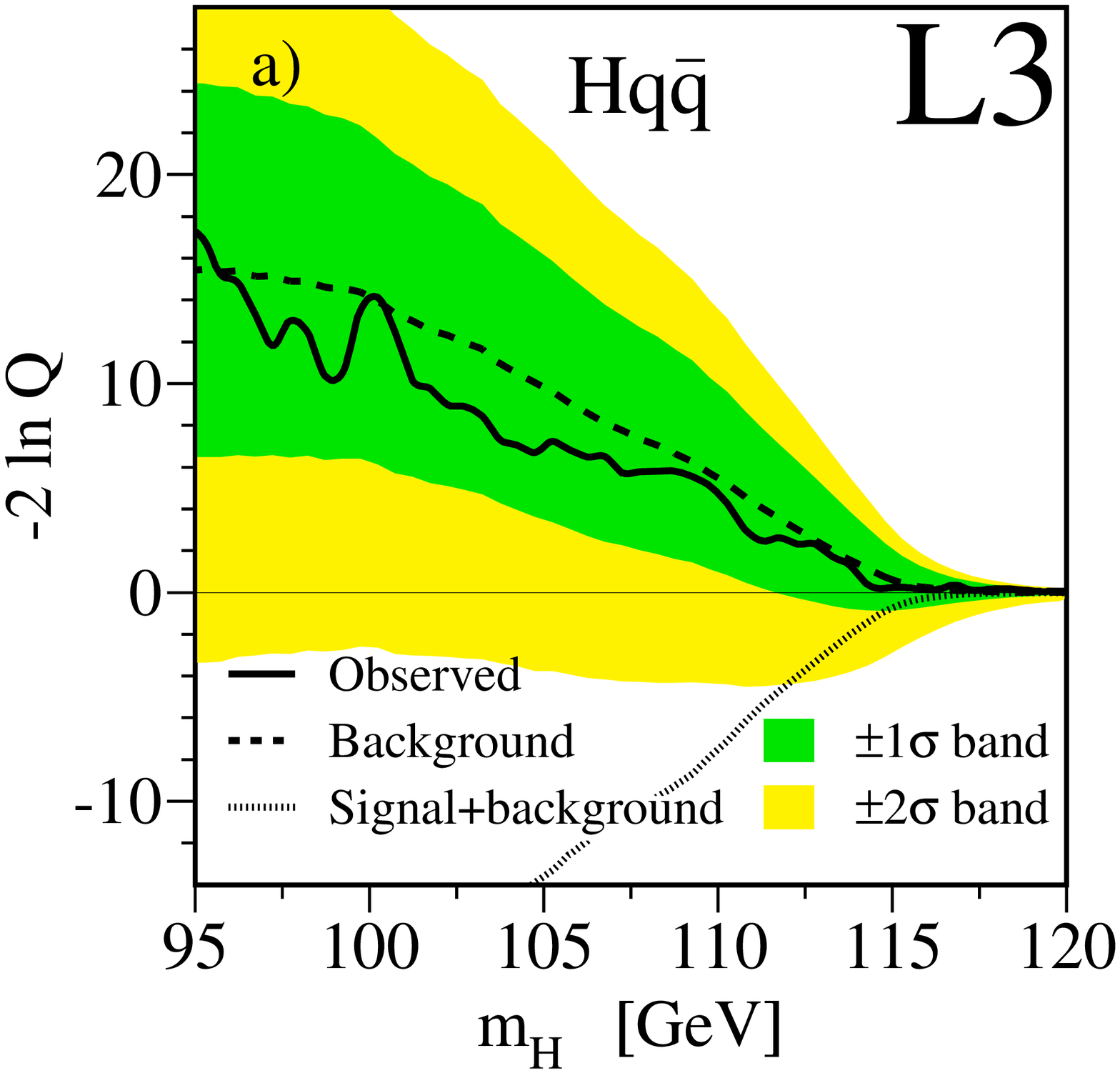,width=0.5\textwidth}%
      \epsfig{figure=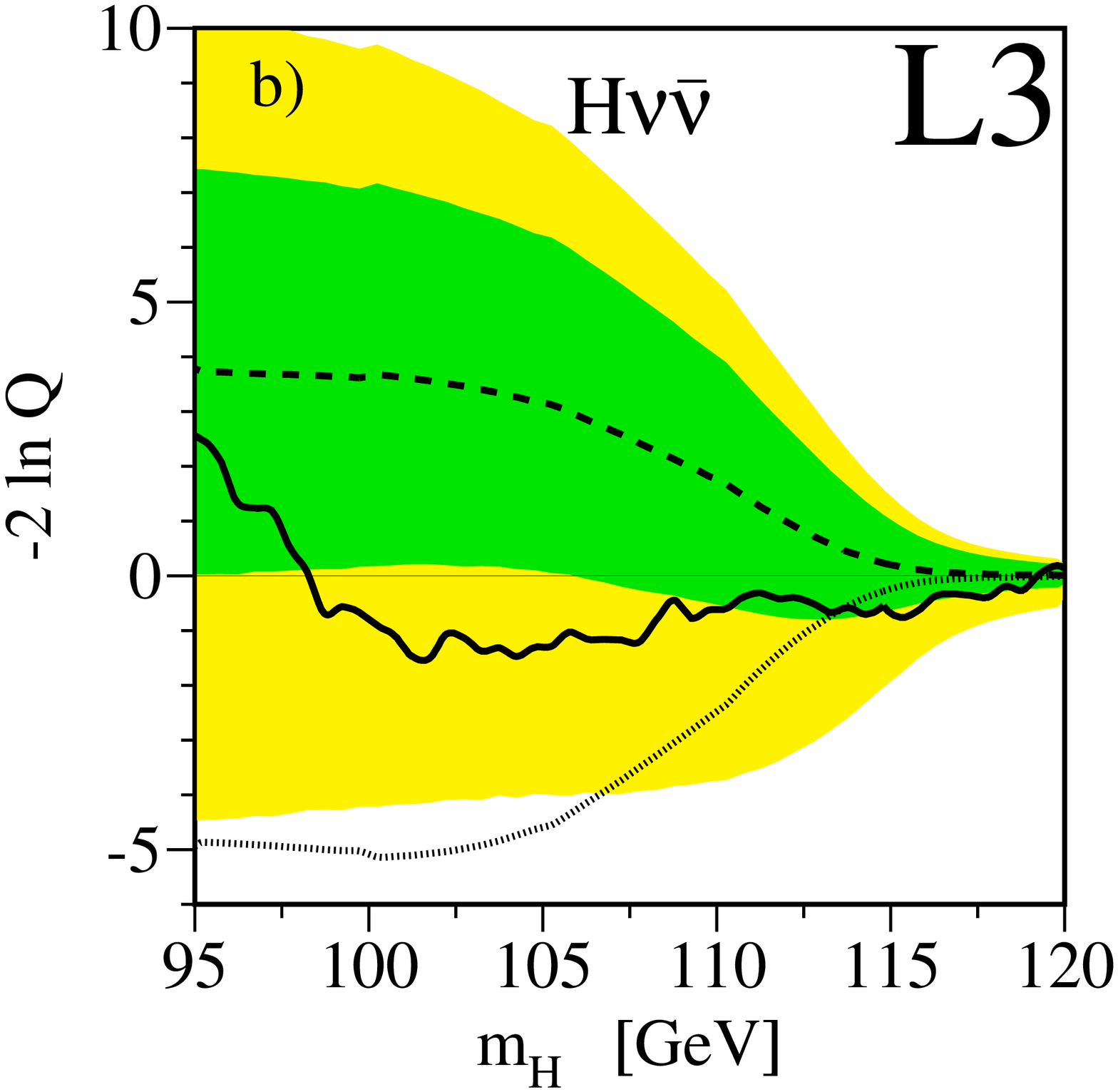,width=0.5\textwidth}}
\mbox{\epsfig{figure=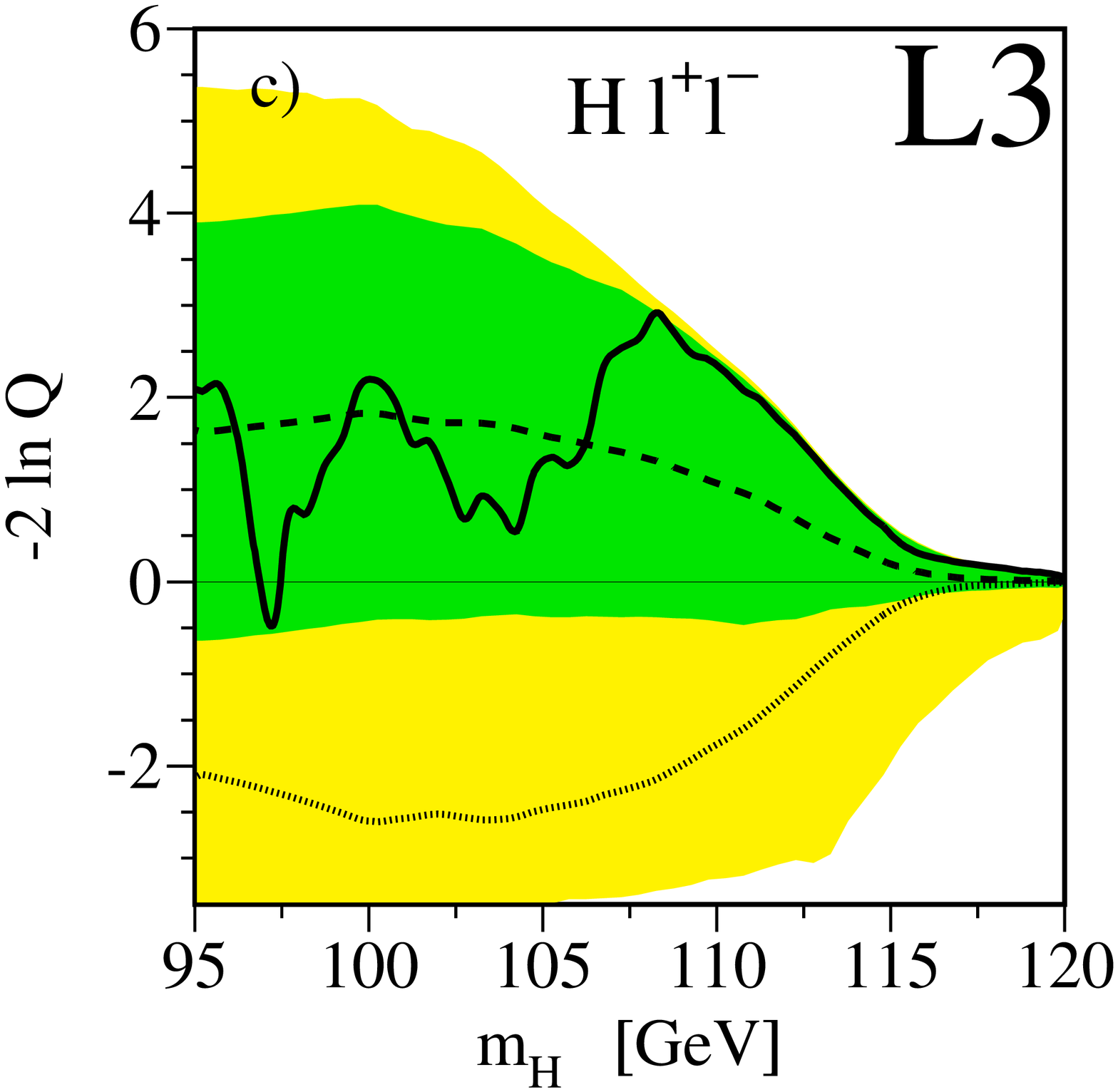,width=0.5\textwidth}%
      \epsfig{figure=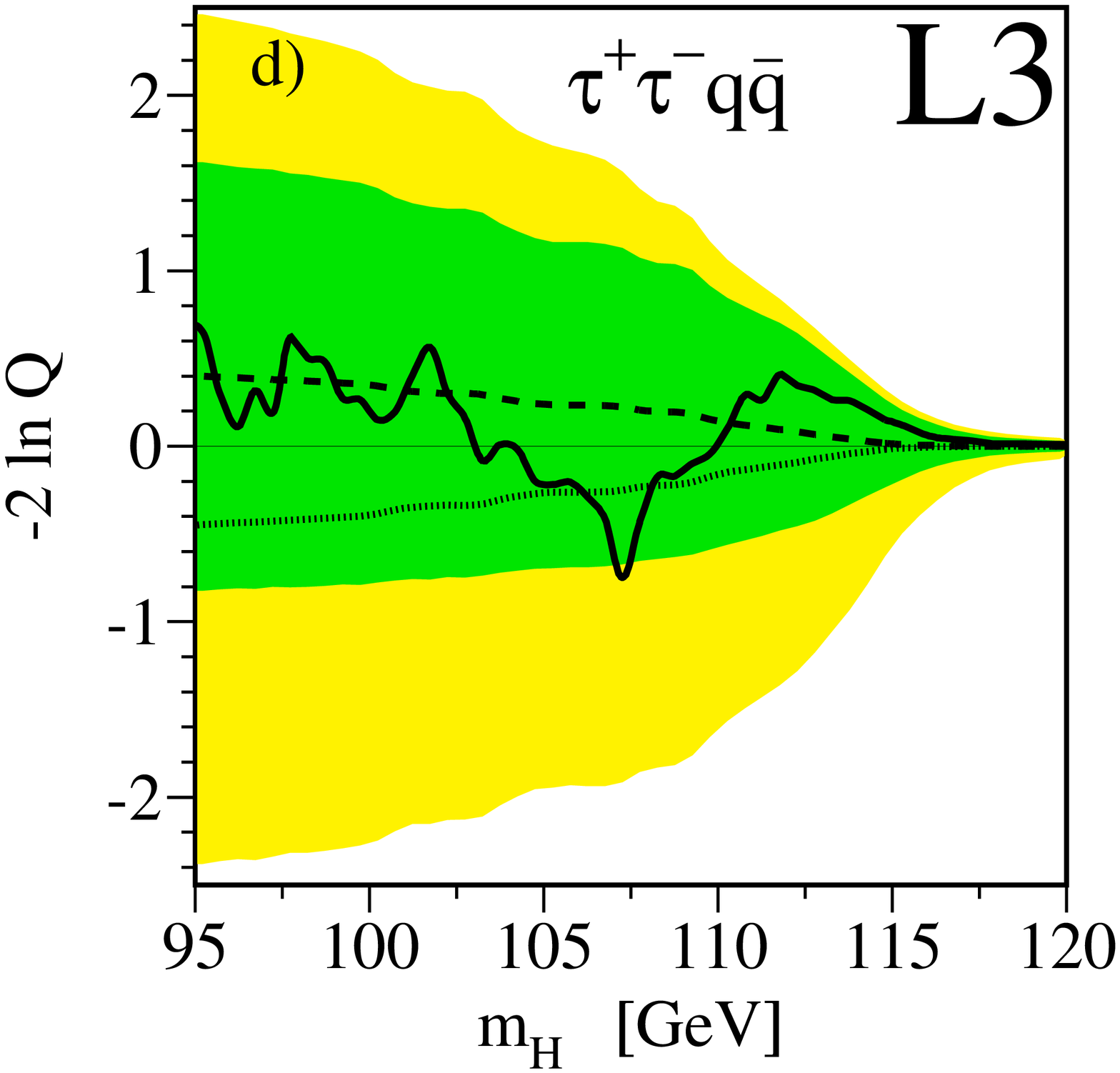,width=0.5\textwidth}}
\caption[]{\label{fig:allllr}  The log-likelihood ratio, $-2 \ln{Q}$, as
           a function of the Higgs mass  hypothesis, \mH, for the search
           channels  a) \Hqq, b) \Hnn, c) \Hll and d) \ttqq.  The  solid
           line shows the observed  $-2  \ln{Q}$.  The dashed line shows
           the   expected   median   value  of  $-2   \ln{Q}$   for  the
           ``background-only''  hypothesis.  The dark and  light  shaded
           bands show the 68\% and 95\% probability intervals centred on
           the background expected median value.  The dotted line is the
           median   expected   value   for   the   ``signal+background''
           hypothesis.}
\end{figure}

\begin{figure}[hp]
\centerline{\epsfig{figure=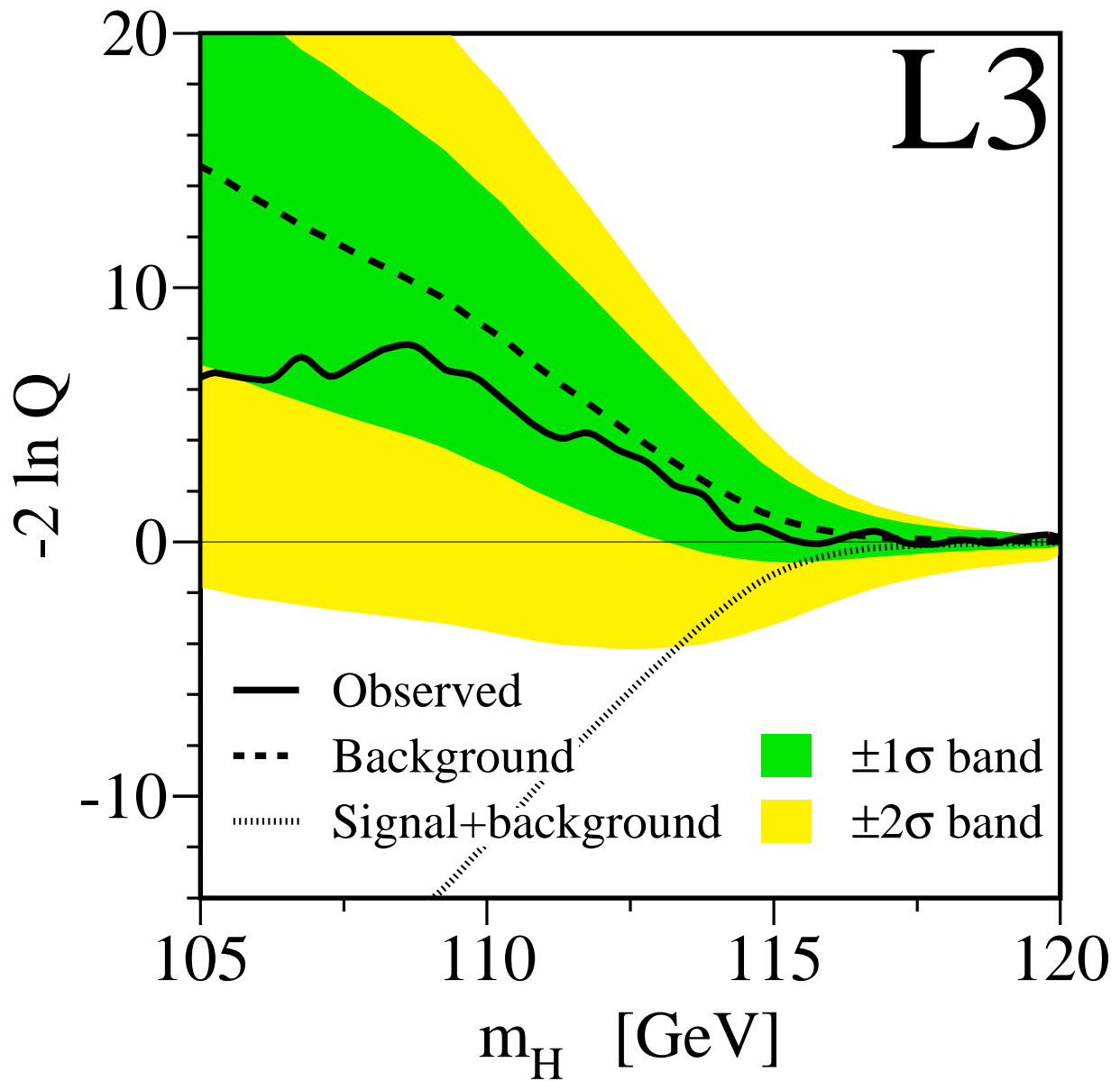,width=0.95\textwidth}}
\caption[]{\label{fig:llrcomb} The log-likelihood ratio, $-2 \ln{Q}$, as
           a function  of the Higgs  mass  hypothesis,  \mH, for all the
           search channels  combined.  The solid line shows the observed
           $-2 \ln{Q}$.  The dashed line shows the expected median value
           of $-2 \ln{Q}$ for the  ``background-only''  hypothesis.  The
           dark  and  light   shaded   bands  show  the  68\%  and  95\%
           probability  intervals  centred  on the  background  expected
           median value.  The dotted line is the median  expected  value
           for the ``signal+background'' hypothesis.}
\end{figure}

Figure~\ref{fig:allllr} shows
the observed $-2\ln{Q}$ compared to the expectation for the 
``background-only'' 
and the ``signal+background'' hypotheses, as a function of \mH, 
for each of the search  channels.
An observed value of   $-2\ln{Q}$ larger than the median expected value 
for the background indicates a deficit of events with respect to the 
expected background while  an observed  $-2\ln{Q}$ value below 
the median expected background value indicates an excess.
Good agreement between the observation and the expected background
is observed in all channels within one standard deviation from the 
background expectation. A slight excess of events above one 
standard deviation from the background is observed in the 
\Hnn channel for \mH above 100\GeV.
The observed and expected log-likelihood ratio $-2\ln{Q}$  
for all channels combined as a function of 
\mH is shown in Figure~\ref{fig:llrcomb}. 

These results are used to evaluate  confidence 
levels for  the ``background-only'' and the ``signal+background'' 
hypotheses. 
The confidence level for the ``background-only'' hypothesis, 
$\CLB$~\cite{lephwg_2000}, is the probability of observing in a large  
sample of simulated ``background-only'' experiments
a more signal-like value of the log-likelihood ratio than is actually 
observed. 
The distribution of  $\CLB$ in a large sample of 
``background-only'' experiments is uniform between 0 and 1, 
thus its median expected value is 0.5. 
An observed value of $\CLB$ lower than 0.5 indicates an excess of events 
in data compared to the expected background.
Similarly, the ``signal+background'' confidence level CL$_{s+b}$ is defined as 
the probability 
in a sample of ``signal+background'' experiments of observing a less 
signal-like value of 
the log-likelihood ratio than  is actually observed. To 
exclude a signal, an additional quantity 
is defined, CL$_s$=CL$_{s+b}$/CL${_b}$~\cite{lephwg_2000}. 
The signal hypothesis is excluded 
at 95\% confidence level when CL$_s$ has a value smaller or equal to 5\%. 
 
The statistical and systematic uncertainties on the signal and 
background expectations are
included in the calculations of the combined confidence levels.
Statistical uncertainties on the
background and signal predictions, 
arising from the finite number of generated Monte Carlo events, 
are evaluated to be up to 8\% for the background and 4\% for the signal. 
The systematic uncertainties are derived using a similar procedure to 
the one adopted in previous Standard Model Higgs
searches~\cite{l3_smh_202}. In addition, 
a systematic uncertainty on the \qqbar background, which affects mostly the 
search region close to the HZ kinematic limit in the \Hnn and \Hqq channels, 
is included depending on \mH. Thus the systematic  uncertainty on the 
number of background events is estimated to be from 6\% up to 15\% 
for \mH close and beyond the HZ kinematic limit. 
The systematic uncertainty on the number of signal events is estimated to
be between 3\% and 6\%, for \mH close to and beyond the HZ kinematic limit,
to take into account the spread  of \come values in the different 
data samples.   

The statistical uncertainty is 
uncorrelated from bin to bin in the final discriminant
distributions and has little effect on the confidence level.  Bins 
of the final discriminant distributions with a
$s/b$ ratio below 0.05 are not considered
in the calculation of the confidence levels, as they degrade the 
search sensitivity once systematic uncertainties are included 
in the calculation.
The number of selected and expected events for all the analyses after such
a $s/b$ cut are summarised in Table~\ref{tab:results}
for the data, the background and the    
Higgs signals for $\mH=110\GeV$ and 115\GeV.  The number of signal
events includes cross-efficiencies from other channels, fusion
processes and charm and gluonic Higgs decays.

\begin{figure}[hp]
\vspace*{-2.cm}
\centerline{\epsfig{figure=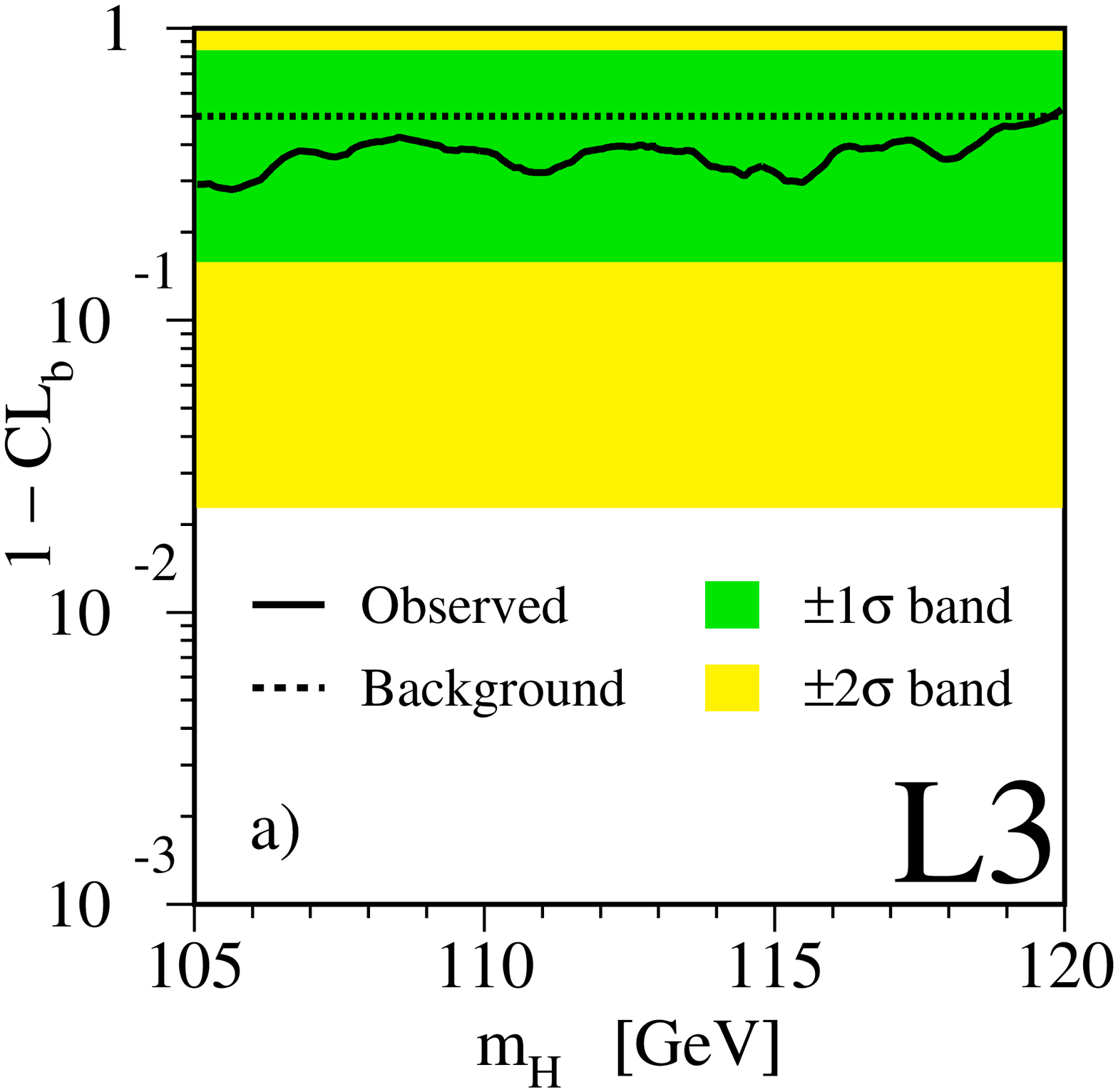,width=0.65\textwidth}}
\centerline{\epsfig{figure=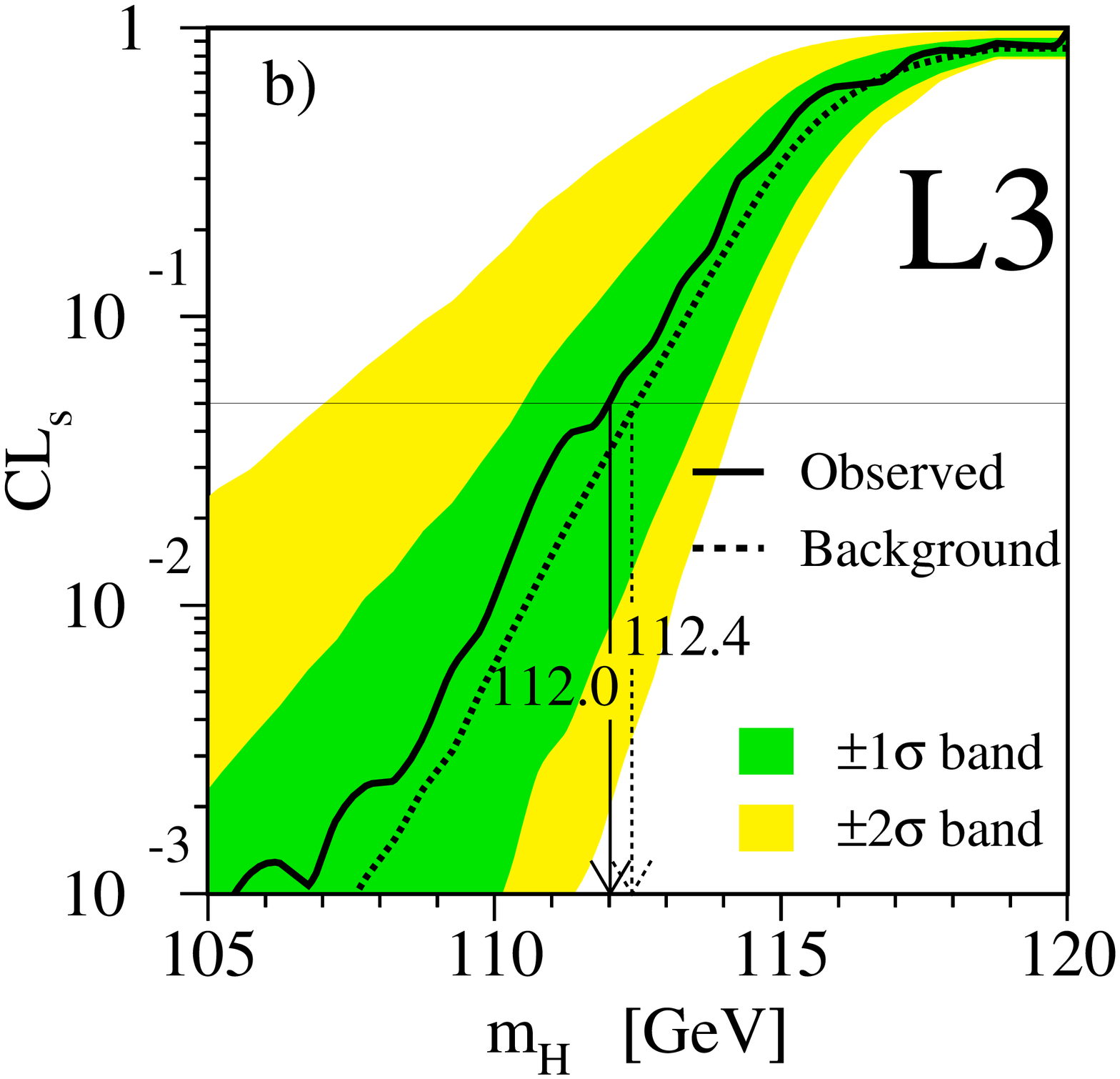,width=0.65\textwidth}}
\vspace*{-0.5cm}
\caption[]{\label{fig:qqalclbcls}  a) The  background  confidence  level
           $\CLB$  and b)~the  signal  confidence  level,  $\CLS$,  as a
           function  of the  Higgs  mass  hypothesis,  \mH,  for all the
           search    channels    combined.   The   data   collected   at
           $189\leq\come\leq   202$\GeV~\cite{l3_smh_189,l3_smh_202}   
           are  also   included   in  the
           combination.  The solid line shows the  observed  value.  The
           dashed line shows the median expected value in a large number
           of simulated  ``background-only''  experiments.  The dark and
           light   shaded  bands  show  the   expected   68\%  and  95\%
           probability  intervals  centred  on the  background  expected
           median value.  The observed  lower limit on the Higgs mass is
           set at 112.0\GeV, with an expected median value of 112.4\GeV,
           at the 95\% confidence level.}
\end{figure}

The confidence level for the ``background-only'' hypothesis 
1-CL$_b$ and the  confidence level for the signal hypothesis CL$_s$ 
as a function of \mH are shown in Figure~\ref{fig:qqalclbcls}. 
They are computed following the procedure of Reference~\citen{lephwg_2000}.
The results of the L3 Standard Model
Higgs searches at lower centre-of-mass
energies~\cite{l3_smh_202,l3_smh_189} are included in
the calculation of these confidence levels.  Values of \mH below 
107\GeV are excluded in the Standard Model 
with  a confidence level greater than 99.5\%.

The observed lower limit on \mH is
112.0\GeV at the  95\% confidence level, for an expected lower limit 
of 112.4\GeV.
This new value improves upon and supersedes our previously
published limit. 
For \mH=112.0\GeV, where CL$_s$ is 5\%, the background probability  
1-CL$_b$ is 40\%. For \mH=115\GeV, 
the background probability is 32\%.
The previously published background probability estimates~\cite{l3_smh_2000_a} 
are consistent 
with the final  results presented here, given the size of the  
uncertainties affecting the 
signal and background estimate in the vicinity of the kinematic limit.


The most significant candidate for \mH=115\GeV is a \Hnn event. 
It has a 
reconstructed Higgs mass of 115\GeV and it was recorded at 
\come=206.4\GeV. 
The kinematic properties of this event were described in detail in 
Reference~\citen{l3_smh_2000_a}.

%
\section*{Acknowledgements}
We acknowledge the efforts of the engineers and technicians who have
participated in the construction and maintenance of L3 and express our
gratitude to the CERN accelerator divisions for the superb performance
of LEP.
%

%

\bibliographystyle{l3style}

%
%
\begin{table}[H]
\begin{center}
\begin{tabular}{|l|ccccccc|}
\hline
\come (GeV)     &  202.8 & 203.8 & 205.1 & 206.3 & 206.6 & 208.0 & 208.6 \\ 
Luminosity (pb$^{-1}$) & 2.7 &  7.6 &  68.1 & 66.9 & 63.7 & 8.2 & 0.1  \\
\hline
\end{tabular}
\caption{The average centre-of-mass energies and
         the corresponding integrated luminosities of the data samples 
collected in the year 2000.}
\label{table:lumi}
\end{center}

\vspace*{0.2cm}

  \begin{center}
    \begin{tabular}{|l|lll|lll|}
      \hline
 & \multicolumn{6}{c|}{Mass hypothesis} \\
\cline{2-7}              
 200$\leq\come\leq$209\GeV& \multicolumn{3}{c|}{$\mH=110\GeV$} & \multicolumn{3}{c|}{$\mH=115\GeV$} \\
\hline
Selection & $N_D$ & $N_B$ & $N_S$ &$N_D$ & $ N_B$ & $ N_S$\\
      \hline
 \Hqq  & 49    & 51.5      & 11.7      & 12     & 9.4     & 1.8      \\
 \Hnn  & 13    & 10.7      &\ 3.3      &\ 5     & 3.3     & 0.66      \\
 \Hee  &\ 0     &\ 0.66     &\ 0.58     &\ 0     & 0.38    & 0.14     \\
 \Hmm  &\ 0     &\ 0.38     &\ 0.45     &\ 0     & 0.26    & 0.11      \\
 \Htt  &\ 0     &\ 0.53     &\ 0.19     &\ 1     & 0.14    & 0.03      \\
 \ttqq &\ 3     &\ 2.3      &\ 0.51     &\ 0     & 0.84    & 0.15      \\
\hline                                                       
 Total & 65    & 66.1      & 16.7     & 18    & 14.3     & 2.9      \\
      \hline
    \end{tabular}
    \caption{The number of observed candidates ($N_D$), 
 expected background ($N_B$) and 
expected signal ($N_S$) events for the data collected in the year 2000, 
after a cut on the final discriminant corresponding to a
      signal-to-background ratio greater than 0.05.  
This cut is used to calculate the confidence levels.}\vspace{0.5em}
    \label{tab:results}
  \end{center}
\vspace*{0.2cm}

\end{table}



~

~

~

~

~

~

~

\newpage

\typeout{   }     
\typeout{Using author list for paper 239 -- ? }
\typeout{$Modified: Jul 3 2001 by smele $}
\typeout{!!!!  This should only be used with document option a4p!!!!}
\typeout{   }
%
%
%
%
%
%

\newcount\tutecount  \tutecount=0
\def\tutenum#1{\global\advance\tutecount by 1 \xdef#1{\the\tutecount}}
\def\tute#1{$^{#1}$}
\tutenum\aachen            
\tutenum\nikhef            
\tutenum\mich              
\tutenum\lapp              
\tutenum\basel             
\tutenum\lsu               
\tutenum\beijing           
\tutenum\berlin            
\tutenum\bologna           
\tutenum\tata              
\tutenum\ne                
\tutenum\bucharest         
\tutenum\budapest          
\tutenum\mit               
\tutenum\panjab            
\tutenum\debrecen          
\tutenum\florence          
\tutenum\cern              
\tutenum\wl                
\tutenum\geneva            
\tutenum\hefei             
\tutenum\lausanne          
\tutenum\lyon              
\tutenum\madrid            
\tutenum\florida           
\tutenum\milan             
\tutenum\moscow            
\tutenum\naples            
\tutenum\cyprus            
\tutenum\nymegen           
\tutenum\caltech           
\tutenum\perugia           
\tutenum\peters            
\tutenum\cmu               
\tutenum\potenza           
\tutenum\prince            
\tutenum\riverside         
\tutenum\rome              
\tutenum\salerno           
\tutenum\ucsd              
\tutenum\sofia             
\tutenum\korea             
\tutenum\utrecht           
\tutenum\purdue            
\tutenum\psinst            
\tutenum\zeuthen           
\tutenum\eth               
\tutenum\hamburg           
\tutenum\taiwan            
\tutenum\tsinghua          

{
\parskip=0pt
\noindent
{\bf The L3 Collaboration:}
\ifx\selectfont\undefined
 \baselineskip=10.8pt
 \baselineskip\baselinestretch\baselineskip
 \normalbaselineskip\baselineskip
 \ixpt
\else
 \fontsize{9}{10.8pt}\selectfont
\fi
\medskip
\tolerance=10000
\hbadness=5000
\raggedright
\hsize=162truemm\hoffset=0mm
\def\r{\rlap,}
\noindent

P.Achard\r\tute\geneva\ 
O.Adriani\r\tute{\florence}\ 
M.Aguilar-Benitez\r\tute\madrid\ 
J.Alcaraz\r\tute{\madrid,\cern}\ 
G.Alemanni\r\tute\lausanne\
J.Allaby\r\tute\cern\
A.Aloisio\r\tute\naples\ 
M.G.Alviggi\r\tute\naples\
H.Anderhub\r\tute\eth\ 
V.P.Andreev\r\tute{\lsu,\peters}\
% ...
\newpage
\begin{list}{A}{\itemsep=0pt plus 0pt minus 0pt\parsep=0pt plus 0pt minus 0pt
                \topsep=0pt plus 0pt minus 0pt}
\item[\aachen]
 I. Physikalisches Institut, RWTH, D-52056 Aachen, FRG$^{\S}$\\
 III. Physikalisches Institut, RWTH, D-52056 Aachen, FRG$^{\S}$
\item[\nikhef] National Institute for High Energy Physics, NIKHEF, 
     and University of Amsterdam, NL-1009 DB Amsterdam, The Netherlands
\item[\mich] University of Michigan, Ann Arbor, MI 48109, USA
\item[\lapp] Laboratoire d'Annecy-le-Vieux de Physique des Particules, 
     LAPP,IN2P3-CNRS, BP 110, F-74941 Annecy-le-Vieux CEDEX, France
\item[\basel] Institute of Physics, University of Basel, CH-4056 Basel,
     Switzerland
\item[\lsu] Louisiana State University, Baton Rouge, LA 70803, USA
\item[\beijing] Institute of High Energy Physics, IHEP, 
  100039 Beijing, China$^{\triangle}$ 
\item[\berlin] Humboldt University, D-10099 Berlin, FRG$^{\S}$
\item[\bologna] University of Bologna and INFN-Sezione di Bologna, 
     I-40126 Bologna, Italy
\item[\tata] Tata Institute of Fundamental Research, Mumbai (Bombay) 400 005, India
\item[\ne] Northeastern University, Boston, MA 02115, USA
\item[\bucharest] Institute of Atomic Physics and University of Bucharest,
     R-76900 Bucharest, Romania
\item[\budapest] Central Research Institute for Physics of the 
     Hungarian Academy of Sciences, H-1525 Budapest 114, Hungary$^{\ddag}$
\item[\mit] Massachusetts Institute of Technology, Cambridge, MA 02139, USA
\item[\panjab] Panjab University, Chandigarh 160 014, India.
\item[\debrecen] KLTE-ATOMKI, H-4010 Debrecen, Hungary$^\P$
\item[\florence] INFN Sezione di Firenze and University of Florence, 
     I-50125 Florence, Italy
\item[\cern] European Laboratory for Particle Physics, CERN, 
     CH-1211 Geneva 23, Switzerland
\item[\wl] World Laboratory, FBLJA  Project, CH-1211 Geneva 23, Switzerland
\item[\geneva] University of Geneva, CH-1211 Geneva 4, Switzerland
\item[\hefei] Chinese University of Science and Technology, USTC,
      Hefei, Anhui 230 029, China$^{\triangle}$
\item[\lausanne] University of Lausanne, CH-1015 Lausanne, Switzerland
\item[\lyon] Institut de Physique Nucl\'eaire de Lyon, 
     IN2P3-CNRS,Universit\'e Claude Bernard, 
     F-69622 Villeurbanne, France
\item[\madrid] Centro de Investigaciones Energ{\'e}ticas, 
     Medioambientales y Tecnolog{\'\i}cas, CIEMAT, E-28040 Madrid,
     Spain${\flat}$ 
\item[\florida] Florida Institute of Technology, Melbourne, FL 32901, USA
\item[\milan] INFN-Sezione di Milano, I-20133 Milan, Italy
\item[\moscow] Institute of Theoretical and Experimental Physics, ITEP, 
     Moscow, Russia
\item[\naples] INFN-Sezione di Napoli and University of Naples, 
     I-80125 Naples, Italy
\item[\cyprus] Department of Physics, University of Cyprus,
     Nicosia, Cyprus
\item[\nymegen] University of Nijmegen and NIKHEF, 
     NL-6525 ED Nijmegen, The Netherlands
\item[\caltech] California Institute of Technology, Pasadena, CA 91125, USA
\item[\perugia] INFN-Sezione di Perugia and Universit\`a Degli 
     Studi di Perugia, I-06100 Perugia, Italy   
\item[\peters] Nuclear Physics Institute, St. Petersburg, Russia
\item[\cmu] Carnegie Mellon University, Pittsburgh, PA 15213, USA
\item[\potenza] INFN-Sezione di Napoli and University of Potenza, 
     I-85100 Potenza, Italy
\item[\prince] Princeton University, Princeton, NJ 08544, USA
\item[\riverside] University of Californa, Riverside, CA 92521, USA
\item[\rome] INFN-Sezione di Roma and University of Rome, ``La Sapienza",
     I-00185 Rome, Italy
\item[\salerno] University and INFN, Salerno, I-84100 Salerno, Italy
\item[\ucsd] University of California, San Diego, CA 92093, USA
\item[\sofia] Bulgarian Academy of Sciences, Central Lab.~of 
     Mechatronics and Instrumentation, BU-1113 Sofia, Bulgaria
\item[\korea]  The Center for High Energy Physics, 
     Kyungpook National University, 702-701 Taegu, Republic of Korea
\item[\utrecht] Utrecht University and NIKHEF, NL-3584 CB Utrecht, 
     The Netherlands
\item[\purdue] Purdue University, West Lafayette, IN 47907, USA
\item[\psinst] Paul Scherrer Institut, PSI, CH-5232 Villigen, Switzerland
\item[\zeuthen] DESY, D-15738 Zeuthen, 
     FRG
\item[\eth] Eidgen\"ossische Technische Hochschule, ETH Z\"urich,
     CH-8093 Z\"urich, Switzerland
\item[\hamburg] University of Hamburg, D-22761 Hamburg, FRG
\item[\taiwan] National Central University, Chung-Li, Taiwan, China
\item[\tsinghua] Department of Physics, National Tsing Hua University,
      Taiwan, China
\item[\S]  Supported by the German Bundesministerium 
        f\"ur Bildung, Wissenschaft, Forschung und Technologie
\item[\ddag] Supported by the Hungarian OTKA fund under contract
numbers T019181, F023259 and T024011.
\item[\P] Also supported by the Hungarian OTKA fund under contract
  number T026178.
\item[$\flat$] Supported also by the Comisi\'on Interministerial de Ciencia y 
        Tecnolog{\'\i}a.
\item[$\sharp$] Also supported by CONICET and Universidad Nacional de La Plata,
        CC 67, 1900 La Plata, Argentina.
\item[$\triangle$] Supported by the National Natural Science
  Foundation of China.
\end{list}
}
\vfill


\newpage
\end{document}